\documentstyle[amsbsy,amsmath,amssymb,psfig,epsf,aps]{revtex}

\def\unity{{\hat{\mathbb I}}}
\def\proj{{\hat{\mathbb P}}}

\title{Structure of Small-Scale Magnetic Fields in the 
Kinematic Dynamo Theory}
\author{Alexander~Schekochihin,\thanks{Present address: 
Imperial College, Blackett Laboratory,
Prince Consort Rd., London~SW7~2BW, U.K.; 
Email address: sure@pppl.gov.}  
Steven Cowley,\thanks{Present address: same as for A.S.;
Email address: steve.cowley@ic.ac.uk.}
Jason Maron,\thanks{Present address: same as for A.S.;
Email address: maron@tapir.caltech.edu.}\\
{\em Department of Physics and Astronomy, 
Box~951547, 
UCLA, Los Angeles, California 90095-1547,}\\
and\\
Leonid Malyshkin,\thanks{Present address: 
University of Chicago, ASCI Flash Center, 5640 S.~Ellis Ave., RI-484, 
Chicago, Illinois 60637-1433; 
Email address: leonmal@flash.uchicago.edu.}\\
{\em Princeton University Observatory, 
Peyton Hall, Princeton, New Jersey 08544}
{\ }}
\date{28 August 2001}

\def\xi{u}

\def\secref#1{Sec.~\ref{#1}}
\def\apref#1{Appendix~\ref{#1}}

\def\exref#1{(\ref{#1})}

\def\eqref#1{Eq.~(\ref{#1})}

\def\figref#1{Fig.~\ref{#1}}

\def\tabref#1{Table~\ref{#1}}

\def\etal{{\it et~al.}}

\def\const{{\rm const}}

\def\bea{\begin{eqnarray}}
\def\eea{\end{eqnarray}}

\def\and{{\quad{\rm and}\quad}}

\def\phi{\varphi}

\def\({\left(}
\def\){\right)}
\def\[{\left[}
\def\]{\right]}
\def\<{\left\langle}
\def\>{\right\rangle}
\def\l{\left}
\def\r{\right}
\def\bl{\bigl}
\def\br{\bigr}

\def\d{\partial}
\def\dt{{\d_t}}
\def\diff{{\rm d}}

\def\vn{\hat{\bf n}}

\def\tZ{{\tilde Z}}

\def\Dt{{{\rm d}\over{\rm d}t}\,}

\def\vx{{\bf x}}
\def\vy{{\bf y}}

\def\vu{{\bf u}}
\def\vB{{\bf B}}
\def\vF{{\bf F}}
\def\vK{{\bf K}}
\def\vM{{\bf M}}
\def\vb{{\skew{-4}\hat{\bf b}}}
\def\vW{{\bf W}}

\def\kpar{k_{\parallel}}
\def\kperp{k_{\perp}}

\def\gpar{\gamma_{\parallel}}
\def\dpar{\nabla_{\parallel}}

\def\kd{k_{\nu}}
\def\kres{k_{\eta}}

\def\Pr{{\rm Pr}}

\def\Bsq{{\langle B^2 \rangle}}
\def\Bfr{{\langle B^4 \rangle}}
\def\Fsq{{\langle F^2 \rangle}}
\def\Ksq{{\langle K^2 \rangle}}
\def\Msq{{\langle M^2 \rangle}}
\def\FB{{\langle F^2/B^4 \rangle}}
\def\FBB{{\langle (\vF\cdot\vB)^2/B^6 \rangle}}
\def\ksq{{\overline{\kpar^2}}}

\begin{document}

\maketitle

\vskip0.1in
\centerline{\tt Published in Phys.~Rev.~E, vol. 65, article 016305 (2002)}

\begin{abstract}


A weak fluctuating magnetic field embedded into a turbulent 
conducting medium grows exponentially while its characteristic 
scale decays. In the interstellar medium and protogalactic plasmas, 
the magnetic Prandtl number is very large, so a broad spectrum of 
growing magnetic fluctuations is excited at small (subviscous) scales. 
The condition for the onset of nonlinear back reaction depends on 
the structure of the field lines. 
We study the statistical correlations that are set up in 
the field pattern and show that the magnetic-field lines 
possess a folding structure, where most of the scale decrease 
is due to the field variation across itself (rapid transverse 
direction reversals), while the scale of the field variation 
along itself stays approximately constant. 
Specifically, we find that, though both the magnetic energy and 
the mean-square curvature of the field lines grow exponentially,  
the field strength and the field-line curvature are anticorrelated, 
i.e., the curved field is relatively weak, while the growing field 
is relatively flat. 
The detailed analysis of the statistics of the curvature 
shows that it possesses a stationary limiting distribution 
with the bulk located at the values of curvature comparable 
to the characteristic wave number of the velocity field 
and a power tail extending to large values of curvature 
where it is eventually cut off by the resistive regularization. 
The regions of large curvature, therefore, occupy only a small 
fraction of the total volume of the system. 
Our theoretical results are corroborated by direct numerical simulations.
The implication of the folding effect is that the advent 
of the Lorentz back reaction occurs when the magnetic energy 
approaches that of the smallest turbulent eddies.
Our results also directly apply to the problem of statistical 
geometry of the material lines in a random flow.\\\\
PACS number(s): 47.27.Gs, 98.35.Eg, 47.65.+a, 05.10.Gg
\end{abstract}

\section{Introduction}

It was demonstrated by Batchelor~\cite{Batchelor_vort_analog} 
that a weak magnetic field passively advected by a turbulent 
velocity field would grow, while its characteristic scale 
would decay. If the magnetic Prandtl number 
(the ratio of fluid viscosity~$\nu$ and magnetic diffusivity~$\eta$, 
$\Pr=\nu/\eta$) 
is large, there is a broad range of subviscous scales available 
to magnetic fluctuations, but not to fluid motions. 
This physical situation is realized in such astrophysical 
environments as the interstellar medium and protogalactic 
plasmas, where $\Pr$~ranges between~$10^{14}$ and~$10^{22}$, 
which provides for~$7$ to~$11$ decades of subviscous range. 
The weak-field (kinematic) regime is believed to represent 
the initial stage of the formation of the currently observed 
magnetic fields of galaxies. These fields, which possess a 
coherent large-scale component and whose energies are comparable 
to the energies of fluid motions of the interstellar 
medium~\cite{Kronberg,Beck_etal,Zweibel_Heiles},  
are thought to have originated from very weak initial 
seed fields in the galaxies (or protogalaxies), which 
have been amplified and brought to their current strength 
and configuration by the dynamo action of the (proto)galactic 
turbulent plasmas (see Refs.~\cite{Kulsrud_review,SBK_review} 
and references therein). 
Constructing a definitive and quantitative theory of this process 
remains an open problem. This theory must necessarily be 
a nonlinear one, because the observed fields are not weak. 
However, developing such a nonlinear theory of the magnetic-field 
evolution will require a thorough understanding of its linear 
(kinematic) precursor. In fact, this point holds with greater 
force in view of the recent theoretical and numerical advances 
which suggest that the saturated spectra of magnetic fluctuations 
are largely determined by the turbulent advection processes 
that drive the kinematic 
dynamo~\cite{Kinney_etal_2D,SMCM_stokes,Maron_Cowley}. 
In this work, we study the geometrical structure of 
the fluctuating small-scale magnetic fields 
produced by the kinematic stage of the high-$\Pr$ dynamo. 
Our findings will have direct bearing on such issues as 
the condition for the onset of nonlinear effects, 
the geometry of the field as it enters the nonlinear stage 
of its evolution, and 
the feasibility of transfering the small-scale magnetic 
fluctuation energy to larger-scale components of the field. 

In an ideal (or highly-conducting) fluid, 
the magnetic fields are (nearly) frozen into the ambient flow.
Therefore, besides being important for the astrophysical dynamo 
as outlined above, studying passive advection of the magnetic field 
is equivalent to studying the statistics of stretching and 
distortion of material lines by random flows, which is 
a fundamental problem in the theory of 
turbulence~\cite{Batchelor_mat_lines}. 
This subject has attracted considerable 
attention~\cite{Batchelor_mat_lines,Cocke_etc,Kraichnan_mat_lines,Drummond_Muench_stretching,Ishihara_Kaneda,Drummond_mat_lines,Pope_curvature,Pope_etal_curvature,Drummond_Muench_curvature,Gluckman_Willaime_Gollub,Muzzio_etal,Boozer_etal}. 
Our results on the statistical geometry of magnetic-field lines 
will have direct applicability in this area. For the sake 
of unity of exposition, we will proceed to develop our 
theory in the language of the kinematic-dynamo problem 
and relegate the drawing of the parallels with the problem of 
material-line advection to the end of the discussion section, 
which concludes this work~(\secref{discussion}).

The mathematical formulation and treatment of the small-scale 
kinematic-dynamo problem were initiated by Kazantsev~\cite{Kazantsev}. 
Kulsrud and Anderson~\cite{KA} developed a detailed spectral 
theory of the small-scale magnetic fluctuations. (A comprehensive 
exposition of the modern state of the second-order statistical 
theory of the small-scale kinematic dynamo with large Prandtl 
numbers, as well as the 
generalization of Kazantsev's and Kulsrud and Anderson's theories 
to the case of arbitrarily compressible velocity fields, can be 
found in~Ref.~\cite{SBK_review}.) 
It was established that the characteristic scale of the advected 
magnetic field decreases exponentially fast at a rate comparable to that 
of the field growth. The magnetic spectrum quickly 
shifts its bulk toward scales extremely small compared to those 
of the velocity field. The decrease of the characteristic scale 
is checked only by the Ohmic resistive dissipation. Such a regime 
persists as long as the kinematic approximation remains valid. 

It is interesting, and, in fact, necessary for a variety of applications, 
to inquire what those small-scale fields 
``look like'': do they really tangle into a completely chaotic 
and fine-scaled web? The most important reason for such an inquiry 
is that it is the structure, not just the strength, of the small-scale 
magnetic fields that determines the conditions for the onset of 
the nonlinear regime. Indeed, we observe that the Lorentz tension 
force~$\vB\cdot\nabla\vB$ only involves the {\em parallel} gradient 
of the magnetic field~\cite{fnote_mag_pressure}. 
Heuristically, the nonlinear Lorentz feedback 
will start to play an important role when the Lorentz tension force 
becomes comparable to inertial terms in the hydrodynamic 
momentum equation, namely, when $B^2\sim (\kd/\kpar)\rho u^2$, 
where $\vu$~is the velocity field, $\kd$~is the smallest-eddy 
wave number, $\rho$~is the density of the medium, and $\kpar$~is 
the characteristic wave number of the magnetic-field variation along itself. 
For chaotically tangled fields, the ratio~$\kd/\kpar$ can be 
as small as $\kd/\kres\sim\Pr^{-1/2}$, where 
$\kres$~is the resistive-regularization wave number. 
The kinematic stage of the dynamo will then only produce very 
weak small-scale fields. On the other hand, if $\kpar$~is restricted 
from growing to be as large as~$\kres$, the kinematic dynamo 
can drive small-scale magnetic fluctuations of energies 
approaching that of the smallest turbulent eddies.   
Much of the previous work on the small-scale dynamo 
and such issues as ambipolar damping and viscous relaxation 
of small-scale magnetic fluctuations was based on specific 
assumptions about the magnitude 
of~$\kpar$~\cite{KA,Chandran_visc_rlx,Subramanian}. 
Understanding the structure of the magnetic field, and, 
in particular, the statistics of the field-line curvature, 
is also crucial for the study of the effect of the Braginskii 
tensor viscosity~\cite{Braginskii} on the small-scale magnetic 
fields~\cite{Kulsrud_etal_report,Malyshkin_thesis}.  

\begin{figure}[t]
\centerline{\psfig{file=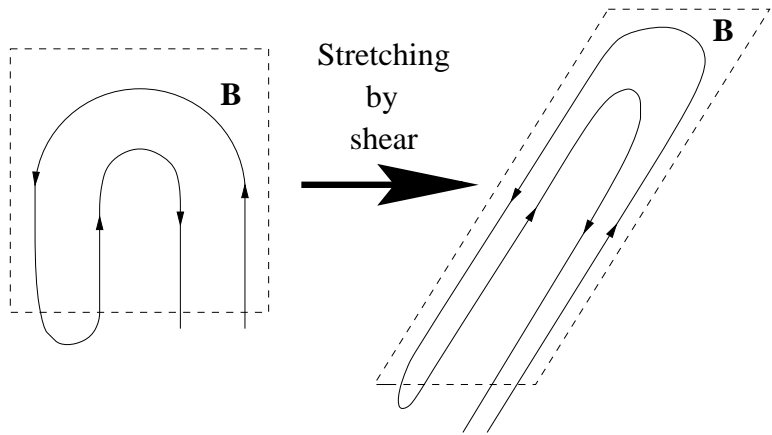,width=4in}} 
\vskip5mm
\caption{Stretching of magnetic-field lines by a linear shear flow.} 
\label{fig_stretch} 
\end{figure}

It was suggested by Cowley~\cite{Cowley_folding} on intuitive grounds 
and later supported by numerical 
simulations~\cite{Kinney_etal_2D,SMCM_stokes,Maron_Cowley} 
that a large-scale 
advecting field, which is locally just a linear shear flow,  
could only stretch the magnetic field and make it flip direction 
ever more rapidly in the plane transverse to the field 
itself~[see~\figref{fig_stretch} and \figref{fig_BK_slices}(a)]. 
It was argued that no appreciable change of the 
characteristic scales at which the magnetic field varies {\em along} 
itself could, therefore, be produced. In other words, the exponential 
increase of the typical fluctuation wave number 
$k = (\kperp^2 + \kpar^2)^{1/2}$ is expected to be due mostly 
to the increase of~$k_\perp$ (rapid transverse direction reversals), 
while $\kpar$ stays approximately unchanged, so $k_\perp\gg\kpar\sim\kd$. 
Such {\em folding nature} of the small-scale fields is consistent 
with the predominance of volume deformations with greatly 
disparate spatial dimensions, which is a well known fact in the 
the theory of kinematic dynamo and passive 
advection~\cite{Chertkov_etal_dynamo,BS_metric}.
It is also, of course, directly related to the extreme 
flux-cancellation property (fine-scale spatial oscillation of 
field orientation) of the dynamo-generated magnetic fields 
in maps and chaotic flows extensively studied by Ott and 
coworkers (see review~\cite{Ott_review} and references therein) 
and by Cattaneo~\cite{Cattaneo_reviews}. 

In this paper, we construct an explicit statistical description 
of the folding effect in the small-scale kinematic-dynamo theory 
and study the correlations that are set up 
between the curvature of the magnetic-field lines and the 
strength of the magnetic field. Since we are interested 
in the geometrical properties of the field, we neglect 
the resistive effects present at extremely small scales 
and consider the diffusion-free induction equation
\bea
\label{B_eq}
\Dt\vB = \vB\cdot\nabla\vu - \vB\nabla\cdot\vu,
\eea
where $\Dt=\dt+\vu\cdot\nabla$ is the full convective 
derivative, $\vB(t,\vx)$~is the passive magnetic field 
and $\vu(t,\vx)$~is the externally prescribed velocity field.  
Let us introduce an auxiliary field~$\vF=\vB\cdot\nabla\vB$, 
which is, of course, the magnetic-tension part of the Lorentz  
force. It is readily seen that~$\vF(t,\vx)$ evolves 
according to the following equation
\bea
\label{F_eq}
\Dt\vF = \vF\cdot\nabla\vu - 2\vF\nabla\cdot\vu 
+ \vB\vB:\nabla\nabla\vu - \vB\vB\cdot\nabla\nabla\cdot\vu.
\eea

Let us first describe a very simple semiquantitative argument that 
supports the folding picture. In the incompressible case 
($\nabla\cdot\vu=0$), we notice that the evolution 
equation~\exref{F_eq} for the Lorentz tension force~$\vF$ 
is identical to that for the magnetic field with the exception of 
the term~$\vB\vB:\nabla\nabla\vu$, 
which contains second derivatives of the velocity field. 
Suppose that an initial distribution of the small-scale magnetic 
fluctuations has been set up in such a way that its characteristic 
parallel and perpendicular wave numbers are comparable and both are 
much greater than the characteristic wave number of the velocity 
field:~$\kpar\sim\kperp\gg\kd$. Then the second derivatives 
of the velocity field can be neglected and the mean-square 
tension force~$\Fsq$ must grow in the same way as the magnetic 
energy~$\Bsq$. For the characteristic wave number of the magnetic 
field we then have
\bea
\label{qualit_arg}
\ksq \sim {\Fsq\over\Bfr} \propto {\Bsq\over\Bfr} 
< \const\,e^{-\gamma_2 t},
\eea 
where $\gamma_2$~is the growth rate of the magnetic energy~$\Bsq$ 
and we have used the obvious fact that~$\Bfr \ge \Bsq^2$.  
Thus, {\em any initial field arrangement where magnetic-field lines 
are chaotically tangled 
will decay toward a folding state at the rate comparable to 
the rate of the magnetic energy growth}~[cf.~\figref{fig_FBK}(a)]. 

In order to see how the situation develops when $\ksq$~becomes 
comparable to~$\kd^2$, a more complete analysis of the statistics 
of the magnetic field and the Lorentz tension is required. 
In~\secref{stats_FB}, we we carry out such an analysis exactly 
for the case of incompressible velocity field, 
and prove that $\ksq=\Fsq/\Bfr$~stabilizes at a value~$\sim\kd^2$.
We then take up the question of the evolution of the magnetic curvature,  
which was recently raised by Malyshkin~\cite{Malyshkin_curvature}. 
We confirm Malyshkin's result on the exponential growth of the 
mean-square curvature. Most importantly, we find that, while 
the ratio of averages~$\Fsq/\Bfr$ tends to a constant, 
the averaged ratio~$\FB$ follows the exponential growth 
of the mean-square curvature. This discrepancy implies 
that the magnetic-field strength and the curvature of 
the magnetic-field lines are very strongly anticorrelated, 
i.e., the magnetic field is weak wherever the curvature 
is large, and vice versa. The picture of folded magnetic-field 
lines is manifestly consistent with this property, while that of 
chaotically tangled ones is not. We argue that the large 
values of curvature in the bends of the folds account for 
the overall growth of the mean-square curvature, even though 
these bends occupy only a small fraction of the total volume 
of the system. At the end of~\secref{stats_FB}, we present a simple 
qualitative description of the folded-magnetic-field-line geometry 
that makes possible the statistical correlations we have found. 

In~\secref{stats_K}, 
we undertake a more detailed study of the one-point distribution 
of the magnetic-field-line curvature and derive equations for its 
probability density function (PDF) 
and all of its moments. This is necessary in order to 
prove the statement of~\secref{stats_FB} that the curvature only 
grows exponentially in a small fraction of the total volume 
of the system. We discover that, while the moments of the curvature 
diverge exponentially in time, its distribution tends to a stationary 
limiting profile whose bulk is concentrated at the values of 
curvature~$\sim\kd$ and which has a power tail at large values of curvature 
(the exponent is~$-13/7$ in the 3D~incompressible 
case). We conclude that the fraction of the volume 
where the growth of the curvature takes place tends to zero with time. 
The limiting values of the curvature moments are determined by the 
resistive regularization at the scales where the magnetic diffusivity 
becomes important:~$\kres\sim\Pr^{1/2}\kd$. 

Our theoretical results on the folding structure of the magnetic field, 
the anticorrelation between the field strength and the field-line 
curvature~(\secref{stats_FB}), the growth of the curvature moments, 
and the stationary limiting distribution of the curvature~(\secref{stats_K}) 
are backed up by the numerical evidence based on the 3D~incompressible 
MHD simulations by Maron and Cowley~\cite{Maron_Cowley}. The relevant 
numerical results are reported at the end of each section.
The agreement between our theory and direct numerical 
simulations of a realistic MHD environment is quite remarkable, 
especially in view of the idealized character of our 
modeling assumptions. 

In~\secref{discussion}, we summarize our findings and 
discuss the implications for the nonlinear dynamo theory. 
The fundamental consequence of the folding effect (i.e., of the fact 
that the parallel scale of the small-scale magnetic fields 
does not decay) is that the nonlinear regime sets in only when 
the magnetic energy becomes comparable to the energy of 
the smallest turbulent eddies. We also explain how our results 
apply to the problem of statistical geometry of material lines 
in isotropic turbulence and relate our conclusions to 
the previous work on this subject. Numerical results obtained 
by several authors in this context provide further confirmation 
of our theory. The fact that the same set of basic features 
of the curvature statistics is found in a number of different 
approaches and models, many of them much more realistic than 
ours, indicates that these statistics may have a largely 
universal character. 

The paper also includes three appendices. 
In~\apref{ap_PDF_derive}, we explain the technical details of 
the derivation of the Fokker-Planck equations used in the paper. 
\apref{ap_compress_effects} is devoted to the study of 
the structure of the small-scale magnetic fields for the case of advecting 
flows that possess an arbitrary degree of compressibility. 
We find that the folding effect as described above 
only persists as long as the degree 
of compressibility of the flow remains below a certain critical value. 
Once this value is exceeded, both the parallel and the perpendicular 
scales of the magnetic-field variation decay exponentially fast 
(albeit at different rates) into the subviscous scale range 
and towards the resistive scales. If this decay continues until 
the parallel and the perpendicular scales are equalized, 
the folding pattern is replaced by the tangled one. 
However, the tangled state is only set up in a small fraction 
of the total volume where the density of the medium is high 
and where most of the magnetic-field growth takes place. 
In the larger (and less dense) part of the system, the magnetic 
field stays relatively weak and flat. 
This new situation brought about by compressibility is due 
to the ability of compressible flows to shrink volumes of the 
medium with frozen-in magnetic-field lines. The structure of the 
field is determined by the competition between stretching and 
contraction. In~\apref{ap_ac_from_metric}, the above consideration 
of the compressibility effects is related to the general 
theory of passive advection in compressible flows developed 
in~Ref.~\cite{BS_metric}.

\section{Statistics of Lorentz Tension and Magnetic-Field-Line Curvature} 
\label{stats_FB}

In this section, we will restrict our consideration to 
the case of incompressible velocity field. 
The evolution equations for the magnetic field~$\vB(t,\vx)$ 
and the Lorentz tension~$\vF(t,\vx)$ in this case are obtained 
from the equations~\exref{B_eq} and~\exref{F_eq} by 
setting~$\nabla\cdot\vu=0$.  
As is customary in the problems of passive 
advection~\cite{Kraichnan_ensemble,Kraichnan_mat_lines} and 
kinematic dynamo~\cite{Kazantsev}, we choose the advecting 
velocity~$\vu(t,\vx)$ to be a Gaussian white-noise-like 
random field ({\em the Kazantsev-Kraichnan flow}), 
whose statistics are defined by its second-order correlation tensor 
\bea
\label{KK_field}
\<\xi^i(t,\vx)\xi^j(t',\vx')\> = \delta(t-t')\kappa^{ij}(\vx-\vx').
\eea
As we will only have to deal with one-point statistical quantities, 
all the relevant information about the velocity correlation properties 
is contained in the Taylor expansion of~$\kappa^{ij}$ 
around the origin~\cite{fnote_Batchelor_regime} 
\bea
\label{xi_024}
\kappa^{ij}(\vy) = \kappa_0\delta^{ij} 
- {1\over 2}\,\kappa_2\bl[y^2\delta^{ij} + 2a y^iy^j\br]
+ {1\over 4}\,\kappa_4 y^2\bl[y^2\delta^{ij} + 2b y^iy^j\br] 
+ \cdots,
\eea 
as~$y\to0$. In order to ensure incompressibility,  
we must set~$a=-1/(d+1)$ and~$b=-2/(d+3)$, 
where $d$~is the dimension of space. Our consideration is formally in 
$d$~dimensions, so that both the two- and the three-dimensional cases 
can be considered in a unified framework. 

The fields $\vB(t,\vx)$ and $\vF(t,\vx)$ satisfy a 
a closed system of equations, 
and, in order to study their statistical properties, we derive 
the Fokker-Planck equation for the joint PDF 
of~$\vB(t,\vx)$ and~$\vF(t,\vx)$ at an arbitrary 
fixed point~$\vx$. Due to the homogeneity of the problem, 
this one-point~PDF~$P(t;\vB,\vF)$ is independent of~$\vx$.
A standard derivation procedure explained in~\apref{ap_PDF_derive} 
leads to the following equation for~$P$:
\bea
\label{FPEq_BF}
\dt P = -{1\over2}\,\kappa^{ij}_{,kl}
\({\d\over\d B^i}B^k + {\d\over\d F^i}F^k\)
\({\d\over\d B^j}B^l + {\d\over\d F^j}F^l\) P
+{1\over2}\,\kappa^{ij}_{,klmn} 
{\d^2\over\d F^i\d F^j} B^k B^l B^m B^n P. 
\eea
The indices following a comma in the subscripts 
always mean spatial derivatives: $_{,k}={\d/\d x^k}$. 
$\kappa^{ij}_{,kl}$ and $\kappa^{ij}_{,klmn}$ are 
the tensors of second and fourth derivatives, respectively, 
of the velocity correlator~$\kappa^{ij}(\vy)$ taken at~$\vy=0$. 
The derivatives with respect to~$B^i$ and~$F^i$ in~\eqref{FPEq_BF} 
act rightwards on {\em all} terms they multiply. 
The Einstein convention of summing over repeated indices 
is used throughout. 
\eqref{FPEq_BF}~contains all the one-point statistical information 
about the distribution of~$\vB$ and~$\vF$ and can, therefore, 
be employed to calculate any individual or mixed averages 
of these quantities. This is done by multiplying~\eqref{FPEq_BF} 
through by the quantity whose average is sought and integrating 
both sides with respect to~$\vB$ and~$\vF$. The derivatives are 
removed via integration by parts and an ordinary differential 
equation is established for the desired average, 
whose time derivative is thereby linked to a linear combination of other 
averages (including itself). The latter averages must in turn be 
calculated in the same fashion. We will see that in many cases of 
interest, very simple linear equations or closed systems of 
linear equations emerge. 

Let us start by calculating the mean-square Lorentz tension. 
We~get   
\bea
\label{Fsq_eq}
\dt\Fsq &=& \gamma_F\Fsq + S_F\Bfr,\\
\label{Bfr_eq}
\dt\Bfr &=& \gamma_4\Bfr.
\eea
The expressions for 
the coefficients~$\gamma_F$, $S_F$, $\gamma_4$, as well as for 
others that will arise in what follows, are collected in~\tabref{tab_gammas}. 
Note that, in accordance with the simple argument we described 
in the Introduction, the growth rate~$\gamma_F$ of~$\Fsq$ is 
the same as that of the magnetic energy~$\Bsq$:~$\gamma_F=\gamma_2$. 

\vskip5mm
{\renewcommand{\baselinestretch}{1.2}
\begin{table}[b]
\begin{center}
\begin{tabular}{c|c|c|c|c|c}
Coefficient & 
Expression &
\multicolumn{2}{c|}{Incompressible} & 
\multicolumn{2}{c}{Irrotational}\\
\hline
 & & $d=3$ & $d=2$ & $d=3$ & $d=2$\\
\hline
\multicolumn{6}{c}{Compressibility parameters}\\
\hline
$a$ & see expansion~\exref{xi_024} & -1/4 & -1/3 & 1 & 1\\
$b$ & see expansion~\exref{xi_024} & -1/3 & -2/5 & 2 & 2\\
$\beta$ & $d[1+(d+1)a]$ & 0 & 0 & 15 & 8\\
$\zeta$ & $d[2+(d+3)b]$ & 0 & 0 & 42 & 24\\
\hline
\multicolumn{6}{c}{Growth rates (from first derivatives of~$\vu$)}\\
\hline
$\gamma_2/\kappa_2$ & 
\parbox{2in}{
$$
{d-1\over d+1}\(d+ 2 + \beta\)
$$} & 5/2 & 4/3 & 10 & 4\\ 
$\gamma_4/\kappa_2$ & 
\parbox{2in}{
$$
2\,{d-1\over d+1}\(d+4+3\beta\)
$$} & 7 & 4 & 52 & 20\\ 
$\gamma_F/\kappa_2$ & 
\parbox{2in}{
$$
{(d-1)(d+2)\over d+1} + {2(3d^2-2)\over d(d+1)}\,\beta
$$} & 5/2 & 4/3 & 65 & 28\\
$\gamma_K/\kappa_2$ & 
\parbox{2in}{
$$
{(9-d)d-2\over d+1} + {2(5-d)\over d(d+1)}\,\beta
$$} & 4 & 4 & 9 & 12\\ 
$\gamma_M/\kappa_2$ & 
\parbox{2in}{
$$
{d-2\over d+1}\(d-1 + {2\beta\over d}\)
$$} & 1/2 & 0 & 3 & 0\\ 
$\gamma_{MK}/\kappa_2$ & 
\parbox{2in}{
$$
{2d\over d+1}\(1+ {\beta\over d^2}\)
$$} & 3/2 & 4/3 & 4 & 4\\ 
\hline
\multicolumn{6}{c}{Source terms (from second derivatives of~$\vu$)}\\
\hline
$S_F/\kappa_4$ & 
\parbox{2in}{
$$
\bl[6(d+4)+\zeta\br]{d-1\over d+3}
$$} & 14 & 36/5 & 28 & 12\\
$S_K/\kappa_4$ & 
\parbox{2in}{
$$
6(d-1)
$$} & 12 & 6 & 12 & 6\\
$S_M/\kappa_4$ & 
\parbox{2in}{
$$
\(6+\zeta\){d-1\over d+3}
$$} & 2 & 6/5 & 16 & 6\\ 
\end{tabular}
\end{center}
\caption{Coefficients for~\secref{stats_FB}. The general formulas 
listed in this table are for the case of arbitrarily compressible flows 
(see~\apref{ap_compress_effects}).
The results for the incompressible case considered 
in~\secref{stats_FB} are obtained by setting~$\beta=0$ and~$\zeta=0$.}
\label{tab_gammas}
\end{table}}

Introducing the characteristic parallel wave number of the magnetic 
fluctuations according to~$\ksq = \Fsq/\Bfr$, we readily find
\bea
\label{ksq_steady}
\ksq(t) = \(\ksq(0)-{S_F\over\gamma_4-\gamma_F}\) e^{-(\gamma_4-\gamma_F)t} 
+ {S_F\over\gamma_4-\gamma_F} 
\to {S_F\over\gamma_4-\gamma_F} 
\sim {\kappa_4\over\kappa_2} \sim \kd^2,\quad t\to\infty, 
\eea
where $\kd$~is the characteristic wave number of the advecting flow. 
The exponential decay of~$\ksq$ 
was already captured in the qualitative argument 
given in the Introduction [see formula~\exref{qualit_arg}]. 
The existence of a steady limiting solution is due to 
the presence of the second derivatives of the velocity 
field in~\eqref{F_eq}. 
By taking them into account, we have thus explicitly proved 
that~$\ksq\sim\kd^2$. 

Let us now undertake a slightly more detailed analysis 
of the magnetic-field structure. 
The Lorentz tension can be decomposed into two 
orthogonal components 
\bea
\vF = B^2\(\vb\cdot\nabla\vb+ \vb\,{\dpar B\over B}\) 
= B^2\(\vK+\vM\),
\eea
where $\vb=\vB/B$~is the unit vector in the direction 
of the magnetic field, and $\dpar=\vb\cdot\nabla$. 
The first term is the magnetic curvature 
vector~$\vK=\vb\cdot\nabla\vb$, the second term, 
$\vM=\vb\dpar B/B$, measures the mirror effect 
and will, for the sake of brevity, be henceforth referred to 
as the mirror force. Since $\vK\perp\vM$, 
we have~$\FB = \Ksq + \Msq$.  
The mean squares of both of these quantities 
can be expressed in terms of mixed averages of~$\vF$ and~$\vB$: 
$\Msq = \FBB$ and $\Ksq = \FB - \FBB$, 
which we proceed to calculate with the aid of~\eqref{FPEq_BF} 
\bea
\label{Ksq_eq}
\dt\Ksq &=& \gamma_K\Ksq + S_K,\\
\label{Msq_eq}
\dt\Msq &=& -\gamma_M\Msq + \gamma_{MK}\Ksq + S_M
\eea
(see~\tabref{tab_gammas} for the values of the coefficients). 
The exact solution of~\eqref{Ksq_eq} is  
\bea
\label{Ksq_sln}
\Ksq(t) = \(\Ksq(0) + {S_K\over\gamma_K}\)e^{\gamma_K t} 
- {S_K\over\gamma_K}, 
\eea
so the magnetic curvature grows exponentially 
(even if it is initially zero). 
It is instructive to express its growth rate 
in terms of the growth rate of the magnetic 
energy~$\Bsq$. In three dimensions, this gives~$\gamma_K=(16/5)\gamma_2/2$, 
which agrees with the result Malyshkin~\cite{Malyshkin_curvature}  
obtained by a direct calculation of~$\Ksq$ in the spirit of 
the Kulsrud-Anderson theory~\cite{KA}. 
We also see that the mean-square mirror force~[\eqref{Msq_eq}] 
is not an independently interesting quantity: 
after a transient initial time, it is reduced to ``mirror'' 
the evolution of the mean-square curvature: 
\bea
\Msq(t) \sim {\gamma_{MK}\over\gamma_K+\gamma_M}\,\Ksq(t), 
\quad t\to\infty. 
\eea

Thus, we have established that, while the ratio of 
the averages~$\Fsq/\Bfr$ tends to a constant value~$\sim\kd^2$, 
the averaged ratio~$\FB\sim\Ksq\sim e^{\gamma_K t}$ grows 
exponentially. Since both of these quantities have the dimension 
and the intuitive meaning of some characteristic parallel 
wave numbers, the question inevitably arises 
as to the physical interpretation 
of such drastic dependence on the relative order of the averaging 
and the normalization with respect to the magnetic-field strength.  
This dependence clearly indicates that 
{\em there exists a very strong anticorrelation between the strength 
of the magnetic field and the curvature of the magnetic-field lines.} 
Namely, while both the mean-square curvature and all moments of~$B$ 
grow exponentially, the magnetic fields are configured 
in such a way that the magnetic field is very weak wherever its 
curvature is large, and vice versa. 
No such arrangement would be possible if the field were 
chaotically tangled everywhere. Indeed, a tangled state 
of this sort would imply 
that the absolute values of the curvature were everywhere 
comparably large and growing. But then, in order to compensate for the 
growth of the curvature, the growth of~$B^4$ 
would have to be partially or fully suppressed compared 
to that mandated by~\eqref{Bfr_eq}. 

\begin{figure}[p]
\centerline{\psfig{file=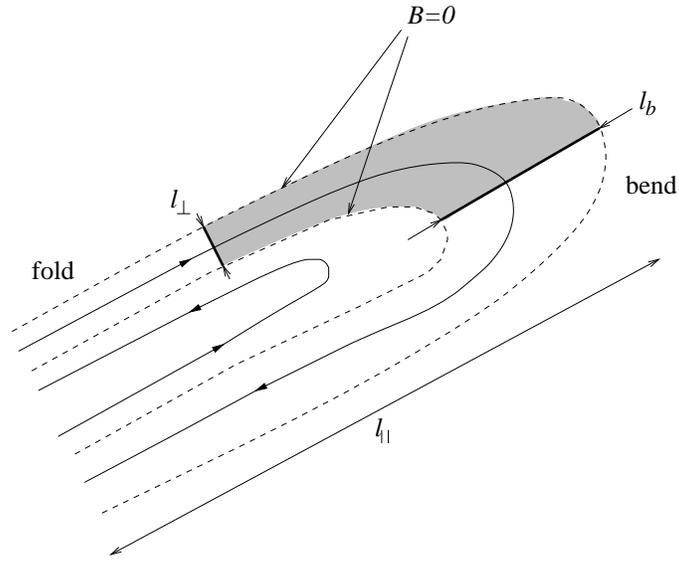,width=3.5in}}
\vskip5mm
\caption{The geometry of the folding field lines in the vicinity 
of the bend.  
The dashed lines correspond to the surfaces on which the magnetic field 
vanishes. The shaded area is the cross section of the volume that can 
be used for the flux-conservation 
estimate~${B_{\rm bend}/B_{\rm fold}}\sim {\ell_\perp/\ell_b}$. 
All the flux is through the surfaces 
whose cross sections are depicted by the bold lines.} 
\label{fig_fold_geom}
\end{figure}

On the other hand, it is easily envisioned how the strong 
anticorrelation between~$B$ and~$K$ can be realized in 
the folding picture. While the curvature is quite small and 
magnetic field grows in most of the volume, which is occupied by 
the folds, the situation is reversed in the small part 
of the volume where magnetic-field lines bend and 
reverse direction: the curvature there is very large and 
magnetic field weak. 
\figref{fig_fold_geom} illustrates the typical geometry 
of the folding magnetic-field lines.  
This picture is in~2D, but can also be interpreted as a cross section 
of a 3D~geometry of a sheetlike configuration. 
Flux conservation ($\int\vB\cdot\diff{\bf S}=0$) implies 
${B_{\rm bend}/B_{\rm fold}}\sim {\ell_\perp/\ell_b}$, where 
$\ell_\perp$ is the characteristic scale of magnetic-field variation 
{\em across} itself in the folding region 
and $\ell_b$ is the characteristic size of the bend. 
The velocity shear that produces (or ``sharpens'') the bend acts 
in such a way that $\ell_\perp$~is decreased while $\ell_b$~is 
amplified, so~$\ell_b\gg\ell_\perp$, whence~$B_{\rm bend}\ll B_{\rm fold}$.

\begin{figure}[p]
$$
\begin{array}{cc}
\psfig{file=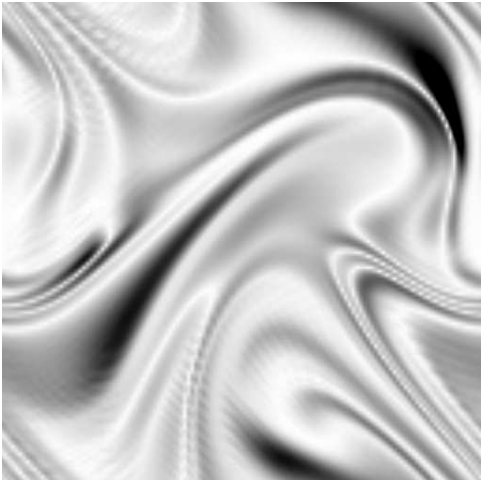,width=2.5in} \qquad & 
\qquad \psfig{file=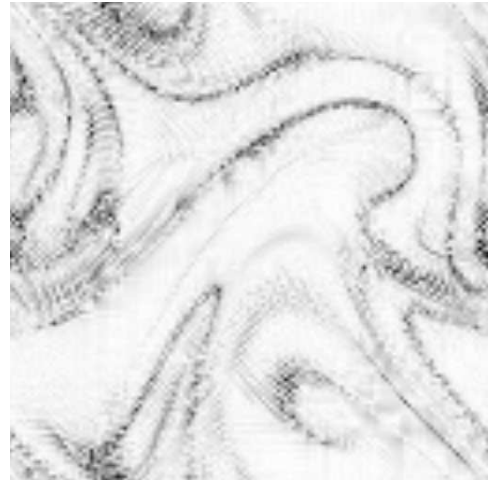,width=2.5in}\\
\qquad & \qquad\\
{\rm (a)~Magnetic~field}~B \qquad & 
\qquad {\rm (b)~Curvature}~K=|\vb\cdot\nabla\vb|
\end{array}
$$
\vskip5mm
\caption{Instanteneous magnetic-field configuration 
in the kinematic regime (numerical results). 
These are greyscale plots of 2D~cross sections of 
the 3D~snapshots of (a)~the magnetic-field strength and 
(b)~the absolute value of the field-line curvature. 
These plots are from the same simulation as~\figref{fig_FBK}(b). 
The field-strength and the curvature snapshots [plots~(a)~and~(b)] 
are taken at the same moment~$t=8.2$ and at the same cross section. 
Darker regions correspond to larger values of the fields. 
In the plot~(a), $\langle B\rangle\simeq 0.003$, 
$\Bsq^{1/2}\simeq 0.004$, 
$\Bfr^{1/4}\simeq 0.006$, 
and the maximum value of~$B$ throughout the system 
is~$\simeq 0.025$. The regions that are pitch black 
in the plot encompass fields stronger than~$0.01$. 
All of these values correspond to magnetic-field energies well 
below the nonlinear-saturation threshold. The specific units 
of the field strength are, of course, of no consequence here.  
In the plot~(b), $\langle K\rangle\simeq 50$, 
$\Ksq^{1/2}\simeq 70$,
$\langle K^4\rangle^{1/4}\simeq 110$, 
and the maximum value of~$K$ is~$\simeq 520$. 
The pitch black regions of  
the plot correspond to curvatures larger than~$400$. 
The curvature has units of inverse length, based on the box size~$1$.}
\label{fig_BK_slices}
\end{figure}

It must be recognized, however, that the presence of an anticorrelation 
between the magnetic-field strength and the magnetic-field-line 
curvature does not in itself prove that the volume where the growth 
of the curvature occurs constitutes only a small fraction 
of the total volume of the system. Indeed, examples of magnetic fields 
can be constructed that possess such an anticorrelation 
and where, at the same time, the mean-square curvature grows 
in any arbitrary fraction of the total volume that can be specified 
beforehand. Further 
study of the curvature statistics is, therefore, required to settle 
this issue. This will be carried out in~\secref{stats_K}, 
where the smallness of the volume where the curvature grows is confirmed. 

Finally, let us reiterate that the presence of the folding structure 
has found repeated confirmation by numerical evidence. 
Most recently, folding was extensively studied in~2D and~3D 
numerical simulations of the small-scale dynamo effect 
in a viscosity-dominated MHD model of Kinney~\etal~\cite{Kinney_etal_2D}, 
and in 3D~forced-MHD simulations of 
Maron and Cowley~\cite{Maron_Cowley}. 
Here we present the numerical results that are based on the latter work 
and directly relate to the theory developed in this 
section. All numerical results presented in this paper 
derive from a $128^3$~spectral forced-MHD code written 
by J.~Maron and described in detail 
in~Refs.~\cite{Maron_Goldreich,Maron_Cowley}. 
The external forcing is on the system-size scale and $\delta$~correlated 
in time. In the simulations quoted in this paper, the hydrodynamic 
Reynolds numbers are quite small, so the advecting fluid flows are 
essentially determined by the balance of the forcing and the viscous 
dissipation. However, this is not really a handicap, 
as the purpose of the numerical results 
presented here is to illustrate the kinematic-dynamo properties 
at subviscous scales. More discussion of this issue can be 
found in~Refs.~\cite{Kinney_etal_2D,SMCM_stokes}. 

\begin{figure}[h]
$$
\begin{array}{cc}
\psfig{file=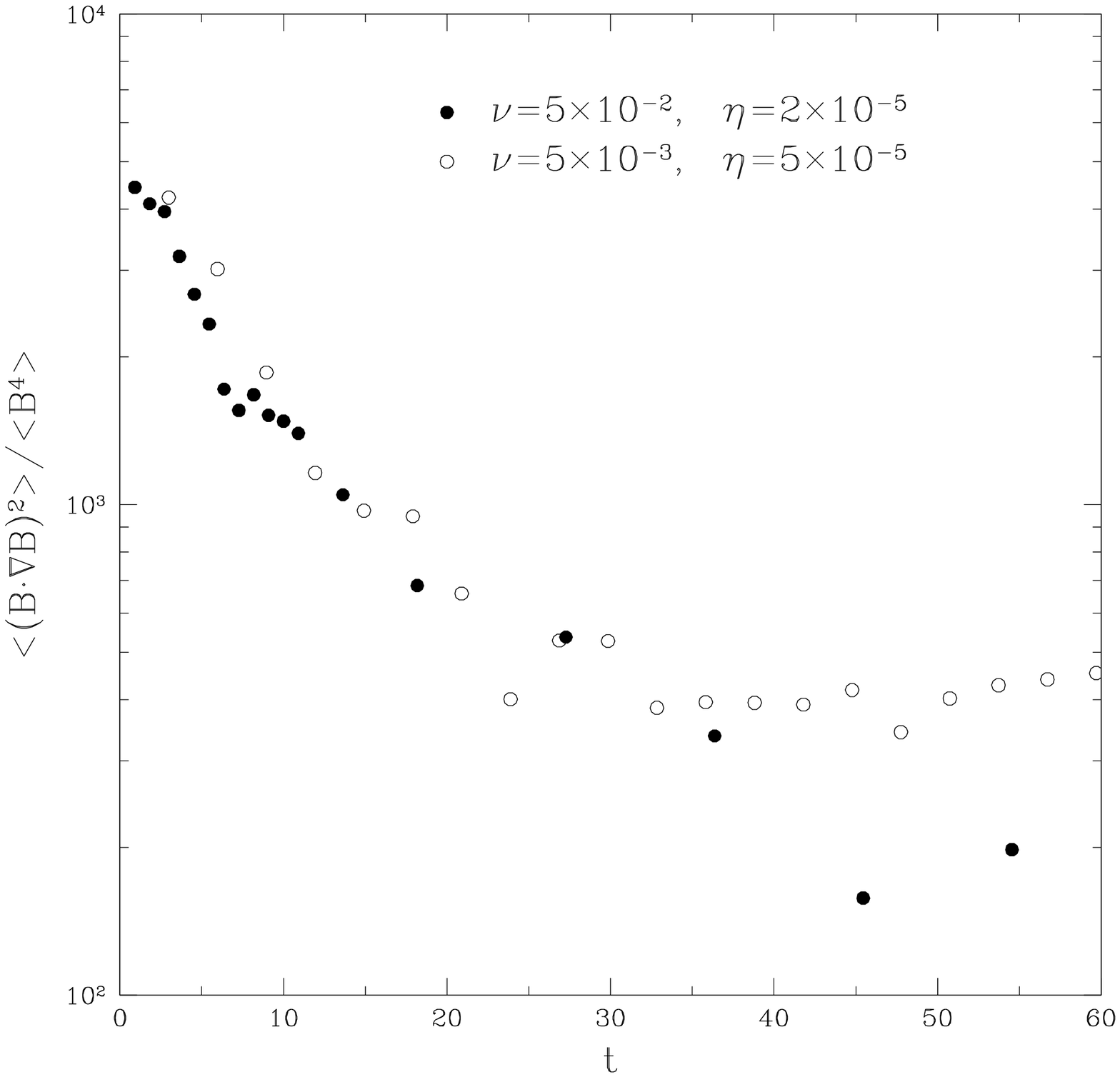,width=3.25in} & 
\psfig{file=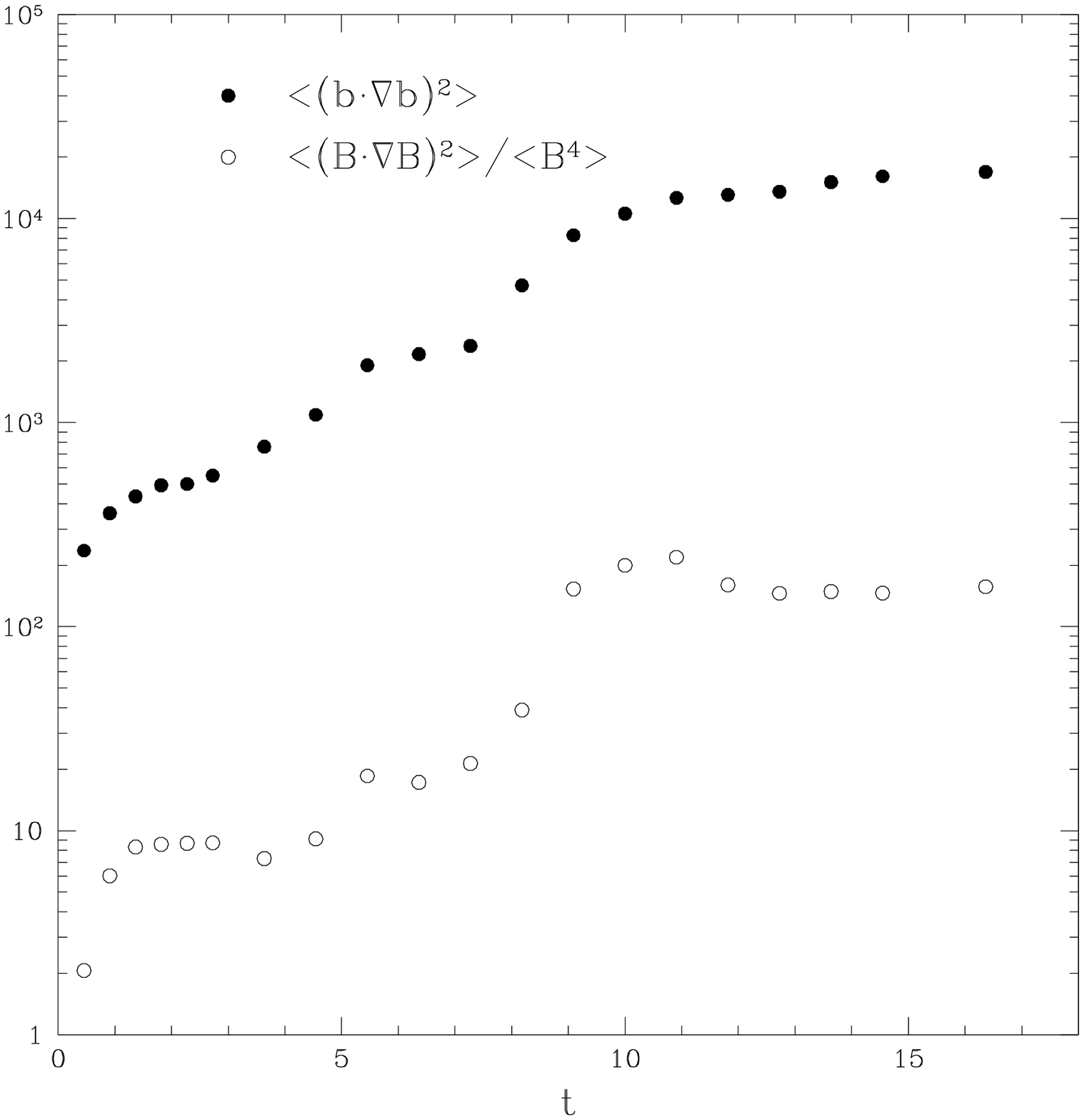,width=3.25in}\\
{\rm (a)~Decay~of}~\Fsq/\Bfr~(\log_{10}~{\rm plot})& 
\qquad{\rm (b)}~\Ksq~{\rm and}~\Fsq/\Bfr~(\log_{10}~{\rm plot})
\end{array}
$$
\vskip5mm
\caption{Anticorrelation between the magnetic-field strength 
and the field-line curvature and growth of the mean-square curvature 
(numerical results). (a)~Time evolution of~$\Fsq/\Bfr$ in two simulations 
where the initial magnetic field is uniformly tangled at subviscous 
scales. The hollow dots correspond to the simulation with~$\Pr=100$, 
$\kd\sim 25$; the filled dots correspond to the simulation 
with~$\Pr=2500$, $\kd\sim 5$. 
(b)~Time evolution of~$\Fsq/\Bfr$ (hollow dots) 
and~$\Ksq$ (filled dots) in a simulation with the initially 
flat magnetic field varying transversely at the velocity scales 
($\kpar=0$, $\kperp\sim\kd$). 
In this simulation, $\Pr=2500$, $\kd\sim 5$, $\kres\sim 250$. 
The ratio~$\Fsq/\Bfr$ again stabilizes at a value~$\sim$~a~few 
times~$\kd^2$. 
In both plots, the quantities plotted have units of inverse length. 
These are based on the box size~$1$. 
Time is measured in units of the smallest-eddy turnover 
time~$\langle|\nabla\times\vu|^2\rangle^{-1/2}$ 
(in units based on box size~$1$ and forcing power~$1$, 
this quantity is~$\sim0.22$).}
\label{fig_FBK}
\end{figure}

In~\figref{fig_BK_slices}, we give 
the greyscale plots of the magnetic-field strength and the absolute 
value of the magnetic curvature corresponding  
to a typical instantaneous magnetic-field configuration 
observed in a simulated MHD evironment 
during the kinematic stage of the small-scale dynamo. 
The folding pattern strikingly similar to the one described above 
is clearly in evidence (cf.~\figref{fig_fold_geom}). 
The anticorrelation between the field strength 
and the field-line curvature, as well as the intermittent 
nature of the distribution of both (cf.~\secref{stats_K}) 
are also manifest. 
\figref{fig_FBK}(a)~shows how  
the ratio~$\ksq=\Fsq/\Bfr$ adjusts to a stationary value~$\sim\kd^2$ 
from an initial state where the field is chaotically 
tangled at subviscous scales. We show results of two simulations 
with such an (artificial) initial field and different values of~$\kd$. 
In both cases, exponential decay of~$\ksq(t)$ toward stationary 
values~$\sim$~a few times~$\kd^2$ is observed, which 
corroborates~\eqref{ksq_steady}. \figref{fig_FBK}(b)~portrays 
the time evolution of the ratio~$\Fsq/\Bfr$ and of the mean-square 
curvature~$\Ksq$ in a simulation that starts with the magnetic field 
concentrated at the velocity scales. The ratio~$\Fsq/\Bfr$ again  
stabilizes at a value~$\sim$~a~few times~$\kd^2$ 
as predicted by our solution~\exref{ksq_steady}. 
The exponential growth of the mean-square curvature~$\Ksq$ 
proceeds in accordance 
with our solution~\exref{Ksq_sln} until it is checked 
by the resistive regularization at a stationary 
value~$\Ksq\sim\kres^2$. While our theory was constructed for 
the diffusion-free regime and, therefore, did not include 
this effect, the resistive saturation of the curvature 
is naturally an expected outcome (see~\secref{stats_K} for more 
discussion of this issue).

\section{Distribution of Magnetic-Field-Line Curvature 
and Magnetic-Field Strength}
\label{stats_K}

In the previous section, we indicated the need for 
a study of the curvature statistics that would go beyond 
the evolution of the mean square. In this section, we fulfill 
this program and delve deeper into the detailed properties of 
the distribution of the magnetic field and its curvature.
 
The Fokker-Planck equation for the one-point~PDF of the 
magnetic-field-line curvature~$\vK=\vb\cdot\nabla\vb$ is most 
conveniently derived 
on the basis of the folowing coupled evolution equations for~$\vK$ 
and the magnetic-field direction~$\vb$:
\bea
\label{K_eq}
\Dt\vK &=& \vK\cdot(\nabla\vu)\cdot(\unity-\vb\vb) 
- \vb\vK\vb:\nabla\vu - 2\vK\vb\vb:\nabla\vu 
+ \vb\vb:(\nabla\nabla\vu)\cdot(\unity - \vb\vb),\\  
\Dt\vb &=& \vb\cdot(\nabla\vu)\cdot(\unity-\vb\vb),
\eea
where $\unity$~is the unit dyadic and colons denote double 
dot products executed according 
to~$\vK\vb:\nabla\vu=\vK\cdot(\vb\cdot\nabla\vu)$,~etc.
Both of the above equations are direct corollaries 
of the induction equation~\exref{B_eq}. 
It is easy to see that these equations respect 
the conservation laws~$|\vb|=1$ and~$\vb\cdot\vK=0$.  
Note that, 
in this section, we work with arbitrarily compressible 
velocity fields, so $\vu$~is not required to be divergence-free. 
It will be seen, however, that none of the essential features 
of the curvature statistics are affected by the compressibility. 

The averaging procedure that leads to the Fokker-Planck equation 
for the joint PDF~$P(t;\vK,\vb)$ does not involve any nonstandard 
steps and is fully analogous to that used to derive the Fokker-Planck 
equation~\exref{FPEq_BF} (see~\apref{ap_PDF_derive}). 
The result is 
\bea
\nonumber
\dt P &=& -{1\over2}\,\kappa^{ij}_{,kl}
\(-\delta^k_i + {\d\over\d b^i}b^k + {\d\over\d K^i}K^k 
- {\d\over\d b^r}b^r b^k b^i - 2{\d\over\d K^r}K^r b^k b^i 
- {\d\over\d K^r}b^r K^k b^i - {\d\over\d K^r}b^r b^k K^i\)\\
\nonumber
&& \quad\quad\,\,\,\times\({\d\over\d b^j}b^l + {\d\over\d K^j}K^l 
- {\d\over\d b^s}b^s b^l b^j - 2{\d\over\d K^s}K^s b^l b^j 
- {\d\over\d K^s}b^s K^l b^j - {\d\over\d K^s}b^s b^l K^j\) P\\ 
&& +\,{1\over2}\,\kappa^{ij}_{,klmn} 
\({\d\over\d K^i}b^k b^m - {\d\over\d K^r}b^r b^k b^m b^i\) 
\({\d\over\d K^j}b^l b^n - {\d\over\d K^s}b^s b^l b^n b^j\) P.
\label{FPEq_Kb}
\eea
A major simplification of this equation becomes possible if one 
recalls that the joint distribution~$P(t;\vK,\vb)$ is subject 
to two constraints: $|\vb|=1$ and $\vb\cdot\vK=0$. 
Also taking into account the spatial isotropy of the problem, 
we conclude that the following factorization holds: 
\bea
\label{ansatz}
P(t;\vK,\vb) = \delta(|\vb|^2-1)\delta(\vb\cdot\vK) P_K(t;K).
\eea
The function~$P_K(t;K)$ is then found to satisfy 
the following reduced Fokker-Planck equation: 
\bea
\nonumber
\dt P_K &=& {1\over2(d+1)}\,\kappa_2
\[\(5d-1 + {6\over d}\,\beta\) K^2 P_K'' 
+ \(11d^2-6d+1 + {2(7d-2)\over d}\,\beta\) K P_K'\r.\\ 
&&\l. + \(d-1\)\(2d-1\)\(3d-1 + {4\over d}\,\beta\) P_K\] 
+ 3\kappa_4\(P_K'' + {d-2\over K}\,P_K'\), 
\label{FPEq_K}
\eea
where primes denote partial derivatives with respect to~$K$, 
the compressibility parameter~$\beta=d[1+(d+1)a]$ is nonnegative 
and vanishes in the incompressible case, 
and $\kappa_2$, $\kappa_4$, and~$a$ are coefficients of the 
small-scale expansion~\exref{xi_024} of the velocity correlator. 
Note that the distribution of the curvature is independent of the second 
compressibility parameter~$b$. 
The normalization rule for~$P_K(t;K)$ 
follows from the normalization of the original~PDF~$P(t;\vK,\vb)$ 
and from the factorization~\exref{ansatz}: 
$(1/2)S_d S_{d-1}\int_0^\infty\diff K\,K^{d-2} P_K(t;K)=1$, 
where~$S_d=2\pi^{d/2}/\Gamma(d/2)$ is the area 
of a unit sphere in $d$~dimensions. Absorbing the geometrical prefactor 
into~$P_K(t;K)$, we conclude that the true~PDF (in the sense that 
it induces a measure on the volume of the system and integrates to unity) 
is~$K^{d-2}P_K(t;K)$. 
Note that, since the curvature vector must always remain 
perpendicular to the direction of the magnetic field, the curvature 
distribution is effectively restricted to~$d-1$ dimensions. 
 
It is now straightforward to establish the set of evolution 
equations for the even moments of the curvature 
\bea
\label{Kn_eq}
\dt\langle K^{2n}\rangle = \[{2(5d-1)\over d+1}\,n - d +
{2\over d}\({6n\over d+1}-1\)\beta\]n\kappa_2\langle K^{2n}\rangle 
+ 6(d+2n-3)n\kappa_4\langle K^{2(n-1)}\rangle,\quad n\ge1
\eea 
For~$n=1$, \eqref{Kn_eq} reproduces the results for the mean-square 
curvature that 
were obtained in~\secref{stats_FB} and~\apref{ap_compress_effects} 
[see~\eqref{Ksq_eq} and \tabref{tab_gammas}]. 
The higher moments of the curvature are coupled to the lower ones in 
a recursive fashion, but also have their own growth rates that increase 
quadratically with~$n$. This latter kind of intermittency is very similar 
to that encountered in earlier studies of the statistics of the 
magnetic-field strength~\cite{Chertkov_etal_dynamo,BS_metric}. 
For the sake of comparison, let us list here the Fokker-Planck equation 
that determines the~PDF~$B^{d-1}P_B(t;B)$ 
of the magnetic-field strength~$B$ 
and the evolution equation for its moments~$\langle B^{2n}\rangle = 
S_d\int_0^\infty\diff B\,B^{d-1+2n} P_B(t;B)$, 
\bea
\label{FPEq_B}
\dt P_B = {1\over2}\,\kappa_2\,{d-1\over d+1}\,
\biggl[\(1+\beta\)B^2 P_B'' 
+ (d+1)(1+2\beta)B P_B' 
+ d(d+1)\beta P_B\biggr],\\
\label{Bn_eq}
\dt\langle B^{2n}\rangle = {d-1\over d+1}\,
\bl[2n+d + (2n-1)\beta\br]n\kappa_2\langle B^{2n}\rangle. 
\qquad\qquad\qquad
\eea
The primes in~\eqref{FPEq_B} denote derivatives with respect to~$B$.
We note that \eqref{Bfr_eq} is a particular case of~\eqref{Bn_eq}. 
Direct derivation of the above equations by averaging the induction 
equation~\exref{B_eq} is quite standard. 
Details can be found in~Ref.~\cite{SK_tcorr}.  
\eqref{FPEq_B}~can also be obtained by integrating out 
the $F^i$ dependence in~\eqref{FPEq_compr} and using the spatial 
isotropy of the magnetic-field distribution. 
\eqref{Bn_eq}~is a direct consequence of~\eqref{FPEq_B}.   

We now turn to the main objective of this section, 
namely, estimating the fraction of the total volume of the system 
where the curvature growth occurs. 
In~\eqref{FPEq_K}, denote by~$D$, $\Sigma$, and~$\Gamma$ 
the coefficients in front of~$K^2 P_K''$, $K P_K'$, and~$P_K$, 
respectively. Now rescale time and curvature according 
to~$D t\Rightarrow t$ and~$K/K_*\Rightarrow K$, 
where~$K_*=(3\kappa_4/D)^{1/2}\sim\kd$ 
(recall that~$\kd$ is the characteristic wave number of 
the advecting velocity field). We can now rewrite~\eqref{FPEq_K} 
in the following nondimensionalized form 
\bea
\label{PK_eq}
\dt P_K = (1+K^2) P_K'' + \(\sigma K + {d-2\over K}\) P_K' 
+ (d-1)(\sigma-d) P_K,
\eea
where we have used the fact that~$\Gamma=(d-1)(\Sigma-Dd)$ 
and denoted
\bea
\label{sigma_def}
\sigma = {\Sigma\over D} 
= {11d^2-6d+1 + 2(7d-2)\beta/d\over 5d-1 + 6\beta/d}. 
\eea
Besides the dimension of space, the only essential 
parameter of the curvature distribution is~$\sigma$, 
which changes with~$d$ and the degree of compressibility. 
The correct boundary conditions for~\eqref{PK_eq} follow 
from the normalizability 
requirement~$\int_0^\infty\diff K\,K^{d-2} P_K(t;K)<\infty$,
\bea
\label{bndry_cdns_K}
\bl[K^{d-2}P_K'(t;K)\br]_{K=0} = 0,\qquad
\bl[K^d P_K'(t;K) + (\sigma-d)K^{d-1}P_K(t;K)\br]_{K\to\infty} = 0.
\eea 

Let us study the evolution of the curvature statistics 
from an initial setting where the curvature is zero 
everywhere:~$K^{d-2}P_K(t=0,K)\propto\delta(K)$. 
While such a $\delta$-like initial distribution is, of course, 
highly artificial, mathematically it is not an anomalous case 
since, as we have seen [\eqref{Kn_eq}], 
the moments of the curvature would grow even from such 
an initial state.
Two distinct asymptotic regimes can be identified in the 
evolution of the curvature distribution.

\underline{Small-curvature regime.} For small values 
of curvature~$K\ll 1$ (i.e., for the dimensional curvature 
much smaller then~$K_*\sim\kd$), \eqref{PK_eq}~reduces 
to what mathematically is a heat equation in~$d-1$ 
dimensions with radial symmetry
\bea
\dt P_K = P_K'' + {d-2\over K}\,P_K'.
\eea
The solution is a heat profile spreading out from the origin 
\bea
\label{PK_heat}
P_K(t;K) = \const\,{e^{-K^2/4t}\over t^{(d-1)/2}}. 
\eea
Multiplying the solution~\exref{PK_heat} by~$K^{d-2}$, 
we find the peak of the~PDF at~$K_{\rm peak}=\sqrt{2(d-2)t}$, 
i.e., it remains at~$K=0$ in~2D and shifts towards larger~$K$ in~3D. 
In either case, the excitation eventually spreads over towards 
larger~$K\sim 1$, where the small-curvature asymptotic regime 
breaks~down.

\underline{Large-curvature regime.}
At large values of curvature~$K\gg 1$, 
the asymptotic form of~\eqref{PK_eq}~is 
\bea
\dt P_K = K^2 P_K'' + \sigma K P_K' + (d-1)(\sigma-d) P_K. 
\eea  
In logarithmic variables, this is a 1D~diffusion equation 
with the drift velocity~$\sigma-1$ 
and with an overall growth rate~$(d-1)(\sigma-d)$. 
The corresponding Green's function~is
\bea
\label{G_lognorm}
G_K(t-t_0;K,K_0) = 
{e^{(d-1)(\sigma-d)(t-t_0)}\over K_0\sqrt{4\pi(t-t_0)}}\,  
\exp\(-{\bl[\ln(K/K_0)+(\sigma-1)(t-t_0)\br]^2\over 4(t-t_0)}\). 
\eea 
Thus, the curvature distribution develops a lognormal tail.
This clearly accounts for the intermittency we have detected in 
the evolution of the curvature moments~[\eqref{Kn_eq}].     
Multiplying the Green's function~\exref{G_lognorm} by~$K^{d-2}$, 
it is not hard to see that the peak of the excitation 
propagates according to 
$K_{\rm peak} = K_0 \exp\bl[(2d-3-\sigma)(t-t_0)\br]$. 
Substituting the value of~$\sigma$~[formula~\exref{sigma_def}], 
we see that~$2d-3-\sigma<0$ in both 
two and three dimensions, so the peak, in fact, propagates {\em backwards} 
towards smaller values of curvature.\\

The conclusion 
from this simple asymptotic analysis is that, after an initial 
transient time, the curvature~PDF should assume the form where 
its bulk is concentrated at the values of (dimensional) curvature 
smaller or comparable to~$K_*\sim \kd$ and a lognormally decaying tail 
is formed at~$K\gg K_*$. The global maximum of the~PDF
is located at~$K=0$ in~2D and at some~$K\sim K_*$ in~3D. 
Thus, in most of the volume of the system, the values 
of the curvature should not greatly exceed~$K_*$. 
The growth of the moments of the curvature is, on the other hand, 
mostly due to the lognormal tail of the distribution. 
Indeed, multiplying the solutions~\exref{PK_heat} and~\exref{G_lognorm} 
by~$K^{d-2+2n}$, we see that the relative importance of the 
small-curvature region decreases, while that of the lognormal 
tail increases. The peak of the function~$K^{d-2+2n}P_K(K)$ for~$n\ge1$ 
always propagates in the {\em forward} direction. 

Of course, once the solution of~\eqref{PK_eq} has cleared 
the region of validity of the small-curvature asymptotic regime, 
a complicated process of probability redistribution is set up.  
As time passes, the lognormal tail gains more weight, 
while the heat profile at small~$K$ spreads out. 
The precise nature of the evolution of the~PDF 
is decided by the interaction between 
the small-curvature (radial-heat) and large-curvature (lognormal) 
regimes in the crossover region~$K\sim K_*$. 
This interaction can affect the {\em entire}~PDF. 
We can gain more insight into what happens by observing 
that \eqref{PK_eq}~has a stationary solution. Indeed, 
let us write~\eqref{PK_eq} in the following explicitly 
conservative form: 
\bea
\label{PK_eq_cons}
\dt P_K = {1\over K^{d-2}}{\d\over\d K} 
K^{d-2}\[(1+K^2)P_K' + (\sigma-d) KP_K\].
\eea
Setting the left-hand side to zero, integrating twice, 
and making use of the boundary conditions~\exref{bndry_cdns_K} 
to eliminate one of the constants of integration, 
we find the following stationary limiting~PDF: 
\bea
\label{PK_st}
K^{d-2}P_K^{\rm (st)}(K) = \const\,{K^{d-2}\over(1 + K^2)^{(\sigma-d)/2}}. 
\eea
This~PDF satisfies the boundary conditions~\exref{bndry_cdns_K}, 
is properly normalizable, and has a power 
tail~$\sim K^{-[\sigma-2(d-1)]}$. 
The values of the exponent~$\sigma-2(d-1)$ 
for the incompressible and irrotational cases in two and three 
dimensions, along with the values of other relevant parameters,  
are collected in~\tabref{tab_coeffs}. 
The curvature distribution can be seen to converge (in the mean-square 
sense) to the stationary profile~\exref{PK_st} 
if we represent the time-dependent solutions 
of~\eqref{PK_eq_cons} in the form~$P_K(t;K)=C(t;K)P_K^{\rm (st)}(K)$ 
and notice that the prefactor~$C(t;K)$ tends to a constant, viz., 
$\dt\bl<C^2\br> = -2\bl<(1+K^2)({\d C/\d K})^2\br>$,
where the averages are with respect to the stationary 
distribution~\exref{PK_st}. 

\begin{figure}[t]
$$
\begin{array}{cc}
\psfig{file=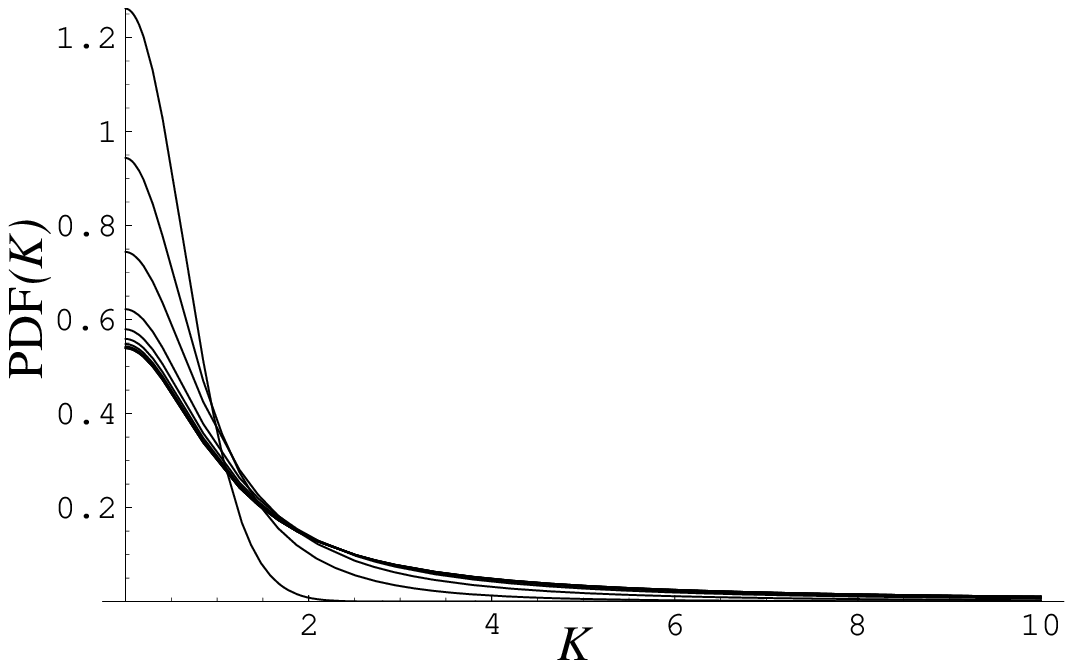,width=3.25in} &
\psfig{file=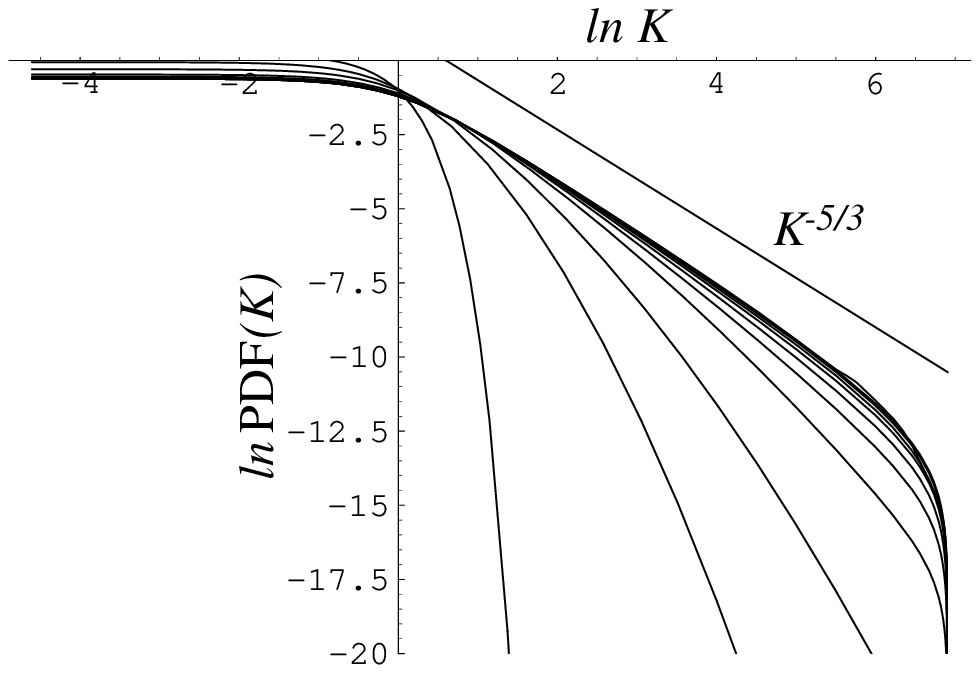,width=3.25in}\\
\qquad & \qquad\\
{\rm (a)}~d=2,~P_K(t;K) & 
{\rm (b)}~d=2,~P_K(t;K),~{\rm ln/ln~plot}\\
\qquad & \qquad\\
\psfig{file=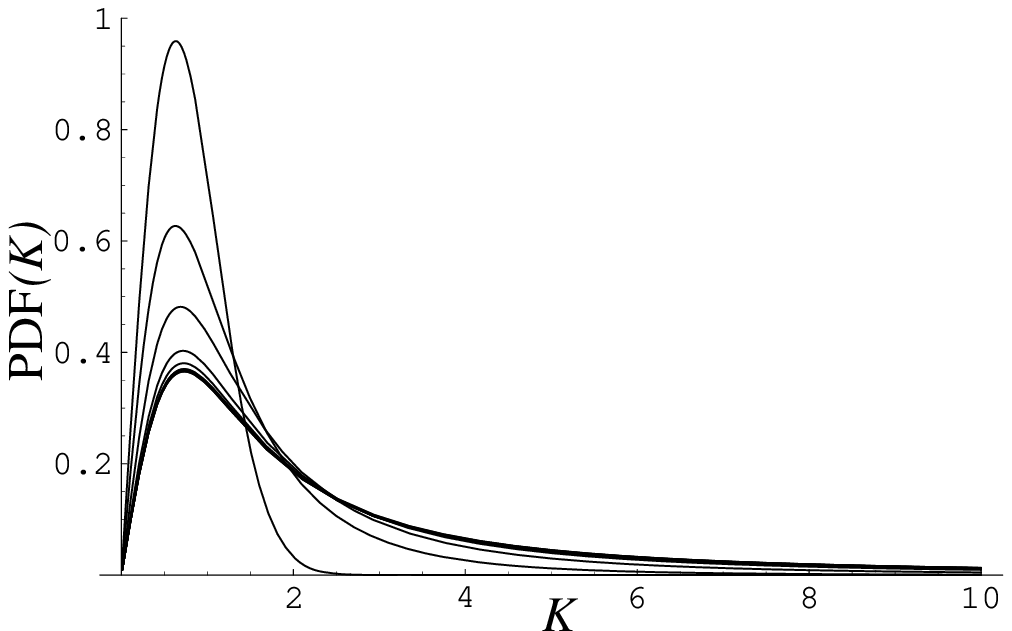,width=3.25in} &
\psfig{file=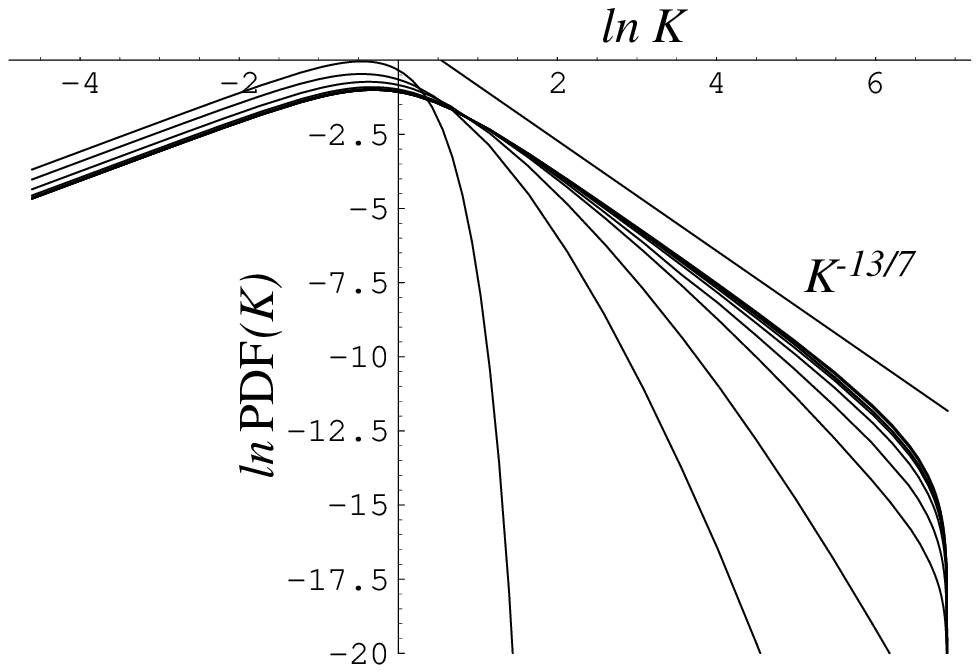,width=3.25in}\\
\qquad & \qquad\\
{\rm (c)}~d=3,~KP_K(t;K) & 
{\rm (d)}~d=3,~KP_K(t;K),~{\rm ln/ln~plot}
\end{array}
$$
\caption{The results of a numerical solution of~\eqref{PK_eq} in two 
and three dimensions for the case of incompressible velocity field.
The numerical solution was initialized with the Gaussian heat 
profile~\exref{PK_heat} corresponding to~$t=0.2$. 
Time is measured in the units of~$D^{-1}$, curvature 
in the units of~$K_*$. 
We plot the evolving~PDF~$K^{d-2}P_K(t;K)$ (normalized to~$1$) 
at times~$t=0.2,0.5,1,2,3,4,5,6,7,8,9$. 
The~PDF that corresponds to the earliest time is 
the one with the highest peak and the steepest decay at large~$K$. 
At later times, the peak of the~PDF descends, while the tail becomes 
thicker (lognormal with increasing variance and eventually powerlike). 
The log/log plots~(b) and~(d) illustrate how the power tail is formed. 
For the sake of reference, we have also plotted 
the slopes correspoding to~$K^{-[\sigma-2(d-1)]}$~(see~\tabref{tab_coeffs}).}
\label{fig_PK}
\end{figure}

\figref{fig_PK}~shows 
the results of numerical solution of~\eqref{PK_eq_cons} 
in two and three dimensions for the case of incompressible 
velocity field. The results for the case of irrotational velocity 
field are very similar in form (with slightly different 
power laws at large~$K$: see~\tabref{tab_coeffs}). 
Collapse of the curvature~PDF 
onto the stationary profile~\exref{PK_st} is very quick and 
proceeds in essentially the same fashion in both incompressible 
and irrotational cases (see discussion at the end of this section). 

\vskip5mm
{\renewcommand{\baselinestretch}{1.2}
\begin{table}[b]
\begin{center}
\begin{tabular}{c|c|c|c|c|c|c|c|c}
$\qquad$ Velocity Field $\qquad$ & 
$\qquad$ Dimension $\qquad$ & 
$\quad a\quad $  & $\quad \beta\quad $ & 
$\quad D/\kappa_2\quad $  & 
$\quad \Sigma/\kappa_2\quad $ & 
$\quad \Gamma/\kappa_2\quad $ & 
$\quad \sigma\quad $ &
$\quad \sigma-2(d-1)\quad $ \\
\hline
Incompressible & $d=3$ & -1/4 & 0  & 7/4  & 41/4 & 10   & 
41/7  & 13/7\\
               & $d=2$ & -1/3 & 0  & 3/2  & 11/2 & 5/2  & 
11/3  & 5/3\\
\hline
Irrotational   & $d=3$ &  1   & 15 & 11/2 & 34   & 35   & 
68/11 & 24/11\\
               & $d=2$ &  1   & 8  & 11/2 & 43/2 & 21/2 & 
43/11 & 21/11\\
\end{tabular}
\end{center}
\caption{The coefficients of~\eqref{FPEq_K} and~\eqref{PK_eq}.} 
\label{tab_coeffs}
\end{table}}

\begin{figure}[p]
$$
\begin{array}{cc}
\psfig{file=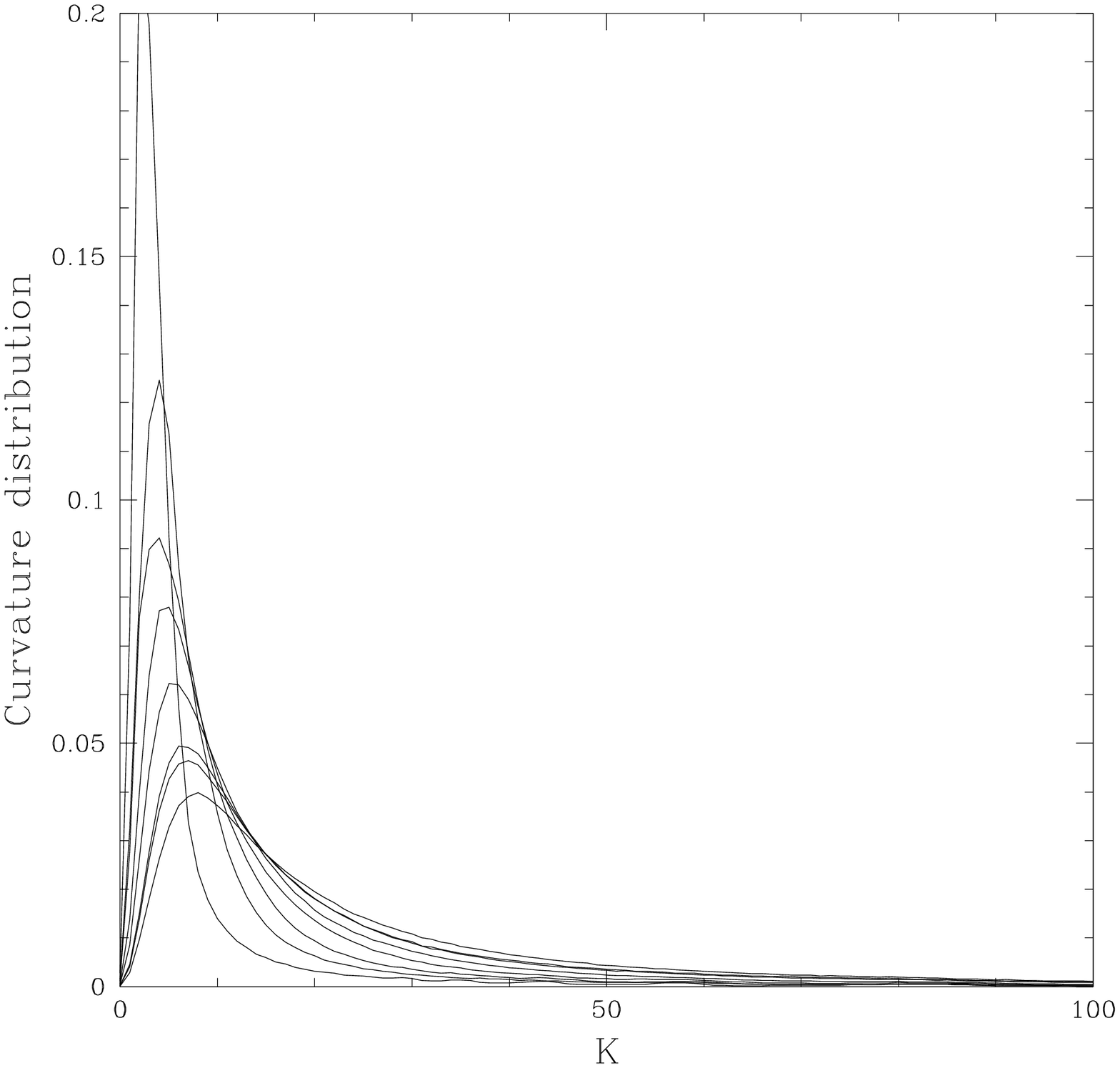,width=3.25in} & 
\psfig{file=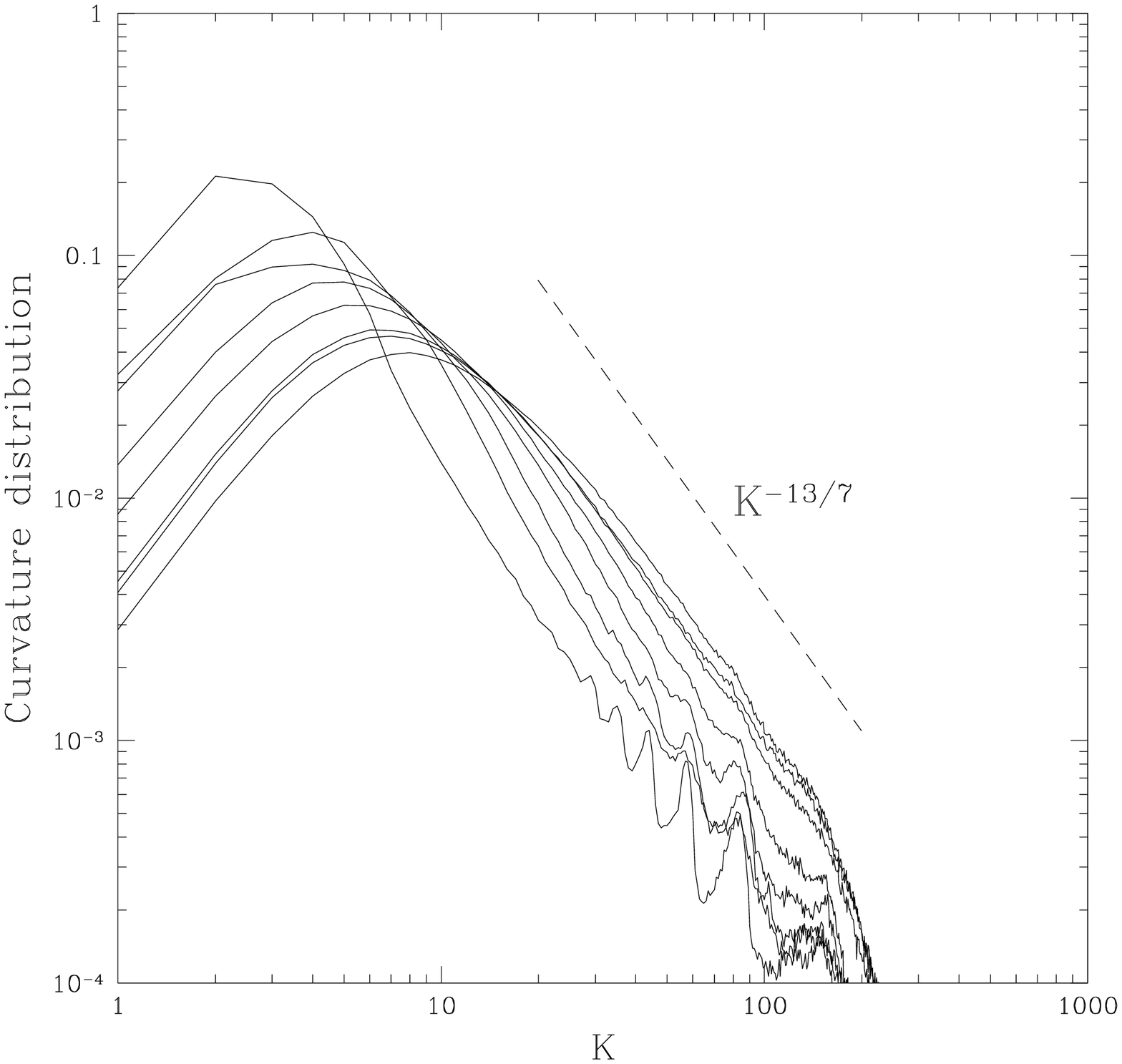,width=3.25in}\\
{\rm (a)}~d=3,~4\pi^2 K P_K(t;K) & 
\quad{\rm (b)}~d=3,~4\pi^2 K P_K(t;K),~{\rm log_{10}/log_{10}~plot} 
\end{array}
$$
\vskip5mm
\caption{Curvature PDF from 3D~incompressible~MHD simulations. 
This is the same simulation as was used in~\figref{fig_FBK}(b): 
$\Pr=2500$, $\kd\sim 5$, $\kres\sim 250$, flat initial field. 
We plot the~PDF~$4\pi^2 KP_K(t;K)$ (normalized to~$1$) 
at times~$t=0.9,1.8,2.7,3.6,4.5,5.5,6.4,7.3$. 
The curvature has units of inverse length, 
based on box size~$1$. The time is measured in the units 
of~$\langle|\nabla\times\vu|^2\rangle^{-1/2}$ 
(the smallest-eddy turnover time).
The~PDF that corresponds to the earliest time 
is the one with the highest peak and the steepest decay at~larger~$K$.
At later times, the peak becomes lower, the decay gentler. 
(a) is the linear plot, (b) is the log/log plot. 
The dashed line represents the slope corresponding to~$K^{-13/7}$.} 
\label{fig_PK_num}
\end{figure}

An important feature of the stationary~PDF~\exref{PK_st} 
is that all the moments~$\langle K^{2n}\rangle$ diverge. 
In the language of physical reality, this means that 
the limiting values of the curvature moments are essentially 
determined by the resistive regularization, which must cut off 
the power tail of the~PDF~\exref{PK_st} at the scale where 
magnetic diffusivity becomes important:~$\kres\sim \Pr^{1/2}\kd$. 
This is, of course, hardly surprising because curvature 
is just a measure of the inverse scale of the magnetic fluctuations 
and cannot exceed the resistive scale. 
In view of these findings, the growth rates for the curvature 
moments that have been obtained in this and the preceding sections, 
should be interpreted as describing the evolution of the moments 
while the lognormal tail of the evolving distribution spreads 
and thickens. The power tail of the stationary limiting 
distribution~\exref{PK_st} forms the envelope 
inside which this process takes place. 
In the diffusion-free regime, the stationary distribution 
itself is attained at~$t\to\infty$ with the moments 
diverging exponentially in time.
We should like to observe here that a~PDF such as we have obtained, 
with a power tail and divergent moments, is indicative of 
a fractal nature of the distribution. 
It must be clear 
that the presence and the particular form of the small-scale 
regularization may affect the global shape of the curvature 
distribution. Since we work in the diffusion-free limit,  
our theoretical results only apply to the period in the evolution 
of the magnetic fluctuations before the small-scale cut off 
is reached. In astrophysical plasmas with very large~$\Pr$, 
this corresponds to an appreciable length of time. In fact, 
current estimates pertaining to the (proto)galactic dynamo 
suggest that the kinematic approximation 
may well break down before the resistive scales become 
important~\cite{Kulsrud_review}.

As regards the distribution of the curvature over the volume 
of the system, the fraction of the volume where the curvature  
exceeds any given value~$K_0$ is easily seen to be 
$V(K>K_0) = \int_{K_0}^\infty\diff K\,K^{d-2}P_K(K)$.
The existence of a stationary distribution 
implies that this quantity tends to a constant that depends on 
the value chosen. Since the bulk of the distribution 
remains at the values of the (dimensional) curvature comparable 
to~$K_*\sim\kd$, the value of~$V(K>K_0)$ for $K_0\gg K_*$ 
will be small. For example, we can use the stationary 
distribution~\exref{PK_st} to estimate that, in the 3D~incompressible 
case, the fraction of the volume where the curvature is more than 
ten~times larger than~$K_*$ does not exceed $14\%$, while the 
fraction of the volume where the curvature is larger than~$100 K_*$ 
is no more than~$2\%$. 
It is, of course, quite clear that {\em the fraction of the volume where 
exponential (or any other kind of) growth of the curvature 
occurs tends to zero with time.} 

Now let us compare the properties of the curvature distribution we have 
just described with the properties of the~PDF of the magnetic-field 
strength determined by~\eqref{FPEq_B}. 
Clearly, the Green's function~$G_B(t-t_0;B,B_0)$ for this equation 
is everywhere lognormal and analogous in form to 
the function~$G_K(t-t_0;K,K_0)$. With time rescaled according 
to~$t\kappa_2(d-1)/2(d+1)\Rightarrow t$, we have
\bea
\label{G_B}
G_B(t-t_0;B,B_0) = 
{e^{d(d+1)\beta (t-t_0)}\over B_0\sqrt{4\pi(1+\beta)(t-t_0)}}\,  
\exp\(-{\bl\{\ln(B/B_0)+\bl[d+(2d+1)\beta\br](t-t_0)\br\}^2\over 
4(1+\beta)(t-t_0)}\). 
\eea 
Multiplying~$G_B(t-t_0;B,B_0)$ by~$B^{d-1}$, we find that 
the peak of the excitation is at 
$B_{\rm peak} = B_0\exp\bl[(d-2-3\beta)(t-t_0)\br]$. 
In the incompressible case, the peak is stationary in~2D and 
propagates {\em forward} in~3D. 
For all compressible flows in~2D and those with~$a>-2/9$ 
in~3D, the direction of propagation is reversed. On the other hand, 
the peak of the function~$B^{d-1+2n}P_B(t;B)$ for~$n\ge 1$ always 
propagates in the forward direction, which accounts for the growth of 
the moments of~$B$~[\eqref{Bn_eq}]. 

Thus, we see that the weakening of the anticorrelation between 
field strength and curvature 
in the compressible flows~(\apref{ap_compress_effects}) 
is due not to any essential change in the properties 
of the curvature distribution, which is quite insensitive 
to the variation of the degree of compressibility of the flow, 
but rather to the fact that the magnetic-field strength itself 
now tends to only grow in a decreasing 
fraction of the total volume of the system. In such a case, 
both the curvature and the magnetic field remain relatively weak 
in most of the volume. An essential difference in their statistics is 
that, unlike the curvature, the magnetic field does not possess 
a stationary limiting distribution. Indeed, due to the scale-invariant 
nature of the Fokker-Planck equation~\exref{FPEq_B}, such a distribution 
would have to be a global power law and hence could not be normalizable. 

In conclusion, we check the main results obtained in this section 
against the numerical evidence supplied by the 3D~incompressible 
MHD simulations of~Maron and Cowley~\cite{Maron_Cowley}. 
In~\figref{fig_PK_num}, we present 
the evolution of the curvature distribution 
observed through a sequence of 
times that correspond to the kinematic and diffusion-free stage of 
these simulations (at later times, the~PDF is affected by the 
resistive regularization and then by the nonlinear effects). 
The collapse of the curvature~PDF onto 
a stationary profile is manifest. The log/log plot 
[\figref{fig_PK_num}(b)] confirms the emergence of the power 
tail~$\sim K^{-13/7}$. 
These results agree very well with our theoretical predictions 
presented in this section.


\section{Summary and Discussion}
\label{discussion}

Let us now summarize the main physical points we have pursued 
in this work and discuss the implications for the nonlinear 
dynamo theory. In this section, we will only discuss the 
case of incompressible flows as the most relevant in the 
astrophysical context we have in mind. The effects of compressibility 
have been given ample attention on both quantitative and 
qualitative level in Appendices~\ref{ap_compress_effects} 
and~\ref{ap_ac_from_metric} and in~\secref{stats_K}.  

In the astrophysical environments that have large magnetic 
Prandtl numbers and, therefore, possess a wide range of 
(subviscous) scales available to the magnetic, but not 
hydrodynamic, fluctuations,  
the small-scale kinematic dynamo is driven by the velocity 
field that locally looks like a linear shear. 
The volume deformations produced by this field lead 
to exponentially fast stretching and folding of the magnetic-field 
lines into a structure characterized by very rapid 
transverse variation of the field, which flips its direction 
at scales ultimately bounded from below only by the resistive length. 
However, the field lines remain largely unbent up to the
scale of the advecting flow. 

Both numerically and analytically, we have established that 
the curvature of the magnetic-field lines and the field strength 
are anticorrelated, i.e., the growth of the field (dynamo)
mostly occurs in the regions of flat field while the sharply 
bent fields remain relatively weak. 
This situation is quickly restored even if the field 
is artificially scrambled into a chaotically tangled state. 
Moreover, in the three-dimensional incompressible flows, 
it is the flat growing fields that occupy most of 
the volume of the system. Accordingly, the field-line 
curvature remains comparable to the inverse velocity scale 
in most of the volume, though its distribution is intermittent 
and all of its moments grow exponentially 
on account of the small regions of strongly bent (but weak) fields. 

In the diffusion-free approximation, i.e., in the regime where 
the magnetic excitaion has not reached the resistive scale,  
the growth of the curvature moments is unbounded and the curvature 
distribution tends to a stationary limiting power-like profile 
with divergent moments. If the resistive cut off is felt 
while the magnetic field is still weak enough to satisfy 
the kinematic assumption, the curvature moments saturate 
at the resistive scale and the precise global shape of the curvature 
distribution may be modified. However, the main features 
of the folding structure described above (the anticorrelation 
between the curvature and the field strength, the smallness of the 
volume where the field is bent) survive because they result
from the large-scale geometric properties of the advection rather 
than from the particular form of the small-scale regularization. 

Let us remark that the reason for some of the statistical 
quantities considered in this paper achieving steady-state 
values even within the confines of the diffusion-free kinematic 
approximation ($\Fsq/\Bfr\to\const$, stationary curvature~PDF) 
is that the second derivatives of the advecting velocity 
field appear in the corresponding dynamic evolution equations 
[see~\eqref{F_eq} and~\eqref{K_eq}]. While passive fields 
such as~$\vB$ only feel the linear component of the ambient velocity 
field and, therefore, have scale-independent nonstationary 
distributions, the statistics of their gradients involve 
an additional scale-dependent parameter~$\kappa_4/\kappa_2\sim\kd^2$. 
In other words, whereas the magnetic field only knows that it is 
advected by a large-scale flow, the statistics of the magnetic-field 
gradients specifically depend on the actual scale size of this 
flow and would not be fully captured in a theory 
where the velocity field were assumed to be linear. 
Physically, the first derivatives of the velocity 
cause the stretching of the field lines (random shear), 
while the second derivatives are responsible for the bending. 

The theoretical results presented in this paper have been 
derived for the Kazantsev-Kraichnan model velocity field~\exref{KK_field}, 
which is Gaussian and $\delta$~correlated in time. 
In view of the highly artificial nature of this field, 
the question naturally 
arises as to the validity of such results in application 
to physical, or even to numerical, realities. 
Indeed, the velocity field that arises in the Kolmogorov turbulence 
is neither Gaussian nor $\delta$~correlated in time: 
its intermittent character is well known and its correlation 
time is naturally estimated to be of the same order as its 
eddy-turnover time. 
However, the Kazantsev-Kraichnan model, while by no means 
a controlled approximation~\cite{SK_tcorr}, 
appears to correctly capture most of the physics of the passive 
advection on at least a semiquantitative level 
and has survived a number of reality tests, 
numerical~\cite{Kinney_etal_2D,SMCM_stokes,Maron_Cowley} 
and, in the case of scalar turbulence, also experimental 
(see, e.g., the review~\cite{Warhaft_review} and references therein). 
Indeed, as was reported in the preceding sections, 
our results are in a very good agreement with numerical simulations, 
where the velocity field derives from the forced Navier-Stokes 
equation and has a realistic correlation time. 

The primary motivation of this study 
of the structure of the magnetic field was  
its crucial importance for the understanding 
of how the nonlinear effects set in. We reserve the detailed 
qualitative and quantitative discussion of this issue for 
an upcoming publication~\cite{SMCM_stokes}. Here we restrict 
ourselves to mentioning the most immediate consequence that the 
folding nature of the small-scale field in the kinematic regime 
has for the onset of the nonlinearity. The Lorentz-feedback 
term in the MHD~momentum equation is proportional to the 
Lorentz tension force~$\vF=\vB\cdot\nabla\vB$~\cite{fnote_mag_pressure}. 
This quantity is quadratic in the magnetic-field strength and 
involves the {\em parallel} gradient of the field. The overall 
effect of the correlations that produce the folding structure is 
to fix the effective value of this parallel gradient at 
approximately the inverse velocity scale~$\kd$. Thus, 
{\em the condition for the nonlinearity to become important 
is the growth of the magnetic energy to values comparable 
to the energy of the smallest turbulent eddies,} rather than 
to much smaller values at which the Lorentz tension 
of a chaotically tangled field would start balancing 
the inertial terms in the momentum equation. 
Any prospects for producing magnetic fluctuations 
at larger scales depend on whether there exists a nonlinear 
mechanism for unwinding the folded structure that the 
nonlinear regime inherits from the kinematic one. 
For further discussion of this subject the reader 
is referred to Refs.~\cite{Kinney_etal_2D,SMCM_stokes,Maron_Cowley}. 

Finally, as was promised in the Introduction, let us 
discuss the relation of our results to 
the fundamental turbulence problem of material-line advection. 
In an ideally  conducting fluid, the behavior  of the magnetic-field 
lines and that of the material lines  are, of course, identical, 
so the diffusion-free kinematic-dynamo problem can be recast 
as a problem of stretching of the material lines by the ambient flow. 
The pioneering work on this subject is due to 
Batchelor~\cite{Batchelor_mat_lines}, who realized that, 
on the average, turbulent motions lead to exponential elongation 
of material line and surface elements. His results were extended 
and, in part, made rigorous by a number of authors~\cite{Cocke_etc,Kraichnan_mat_lines,Drummond_Muench_stretching,Ishihara_Kaneda,Drummond_mat_lines}. 
Starting with the work of Pope~\cite{Pope_curvature}, 
much attention was focused on the statistics of the extrinsic principal 
curvature of material surface elements in turbulent flows~\cite{Pope_curvature,Pope_etal_curvature,Drummond_Muench_curvature,Gluckman_Willaime_Gollub} 
and of the curvature of material lines in both 
turbulent~\cite{Drummond_Muench_curvature,Ishihara_Kaneda,Drummond_mat_lines} 
and deterministic but chaotic~\cite{Muzzio_etal} flows.
Our results on the statistics of magnetic-field-line curvature 
are subject to direct comparison with these earlier studies. 
Indeed, our~\eqref{K_eq} for the vector 
curvature~$\vK=\vb\cdot\nabla\vb$ of the magnetic field line 
leads to the following equation for~$K=|\vK|$: 
\bea
\label{DM_eq}
\Dt K = -\bl(2\vb\vb:\nabla\vu - \vn\vn:\nabla\vu\br) K 
+ \vb\vb:(\nabla\nabla\vu)\cdot\vn,
\eea
where $\vn=\vK/K$~is the unit normal to the field line.
\eqref{DM_eq}~is the same as the equation derived by 
Drummond and M\"unch~\cite{Drummond_Muench_curvature} 
for the curvature of the material line elements (in the usual 
geometric definition) and is very similar (though not 
identical) to Pope's~\cite{Pope_curvature} equation for 
the extrinsic principal curvature of the material 
surface elements. Formal similarity between the curvature 
equations for material lines and surfaces led 
Drummond and M\"unch~\cite{Drummond_Muench_curvature} 
to conjecture that, in isotropic turbulence, the general 
features of the curvature statistics for these objects would 
also be alike~\cite{fnote_mat_surf}. All of the extant numerical evidence 
supports this conjecture at least on the qualitative 
level~\cite{Pope_etal_curvature,Drummond_Muench_curvature,Ishihara_Kaneda,Drummond_mat_lines,Muzzio_etal}. The curvature statistics 
also appear to be largely insensitive to the type of flows 
considered. Thus, numerical simulations involving 
3D~forced~\cite{Pope_etal_curvature} and 
2D~decaying~\cite{Ishihara_Kaneda} Navier-Stokes turbulence, 
Kraichnan's~\cite{Kraichnan_random_flow} random flow 
model~\cite{Drummond_Muench_curvature,Drummond_mat_lines}, 
2D~and 3D~deterministic but chaotic flows~\cite{Muzzio_etal}, 
and, finally, our own 3D~forced-MHD simulations and theoretical 
results based on the Kazantsev-Kraichan velocity field 
(both compressible and incompressible), 
consistently reveal the same set of properties of the curvature 
distribution. The unbounded exponential growth of the mean-square 
curvature was first observed in the numerical studies 
of Pope and coworkers~\cite{Pope_etal_curvature}, who also 
found that, unlike the moments of the curvature itself, 
the moments of its logarithm tended to time-independent asymptotic values. 
The emergence of a stationary curvature~PDF with a power 
tail was reported. Drummond~\cite{Drummond_mat_lines} confirmed 
these results in his numerical model and also offered a qualitative 
theoretical argument that related the existence of the power tail 
of the curvature~PDF to the competition between the stretching 
action of the first spatial derivatives of the velocity field 
and the bending effect of its second derivatives. 
Ishihara and Kaneda~\cite{Ishihara_Kaneda} attempted to determine 
the power law (in~2D) by looking for a critical index~$p_c$ 
such that the curvature moments of orders higher than~$p_c$ 
would diverge. For a model velocity field essentially equivalent 
to the Kazantsev-Kraichnan flow, they deduced~$p_c=2/3$, which was 
consistent with their numerical results for a realistic 
(decaying) 2D~turbulence. All of the above is in perfect agreement 
with the results of~\secref{stats_K} (in particular, 
Ishihara and Kaneda's~\cite{Ishihara_Kaneda} 
critical index exactly corresponds to our $-5/3$~power tail in~$d=2$). 
Drummond and M\"unch~\cite{Drummond_Muench_curvature} also 
(numerically) measured the correlation between curvature and 
stretching and found it negative, as did we in our theory 
and simulations~(\secref{stats_FB}). Such anticorrelation 
was also noticed by Boozer and coworkers~\cite{Boozer_etal}, 
who, in their theory 
of finite-time Lyapunov exponents for chaotic flows, found that 
the local Lyapunov exponent of the flow was strongly suppressed 
in the regions of high curvature. 
Finally, results very similar to those surveyed above were 
obtained by Muzzio and coworkers~\cite{Muzzio_etal} for 
a number of 2D~and 3D~deterministic chaotic flows. 
Thus, the exact theory of the curvature statistics 
that we have been able to develop in the framework of the 
Kazantsev-Kraichnan model incorporates all of the essential features 
thus far observed numerically, as well as surmised in less 
direct theoretical ways. The close agreement between our theory 
and an array of numerical results obtained for more realistic 
flows provides an additional validation of our approach 
and, more importantly, suggests that the statistics of line- 
and surface-element advection possess a high degree of universality.

\section*{Acknowledgments}

It is a pleasure to thank Russell~Kulsrud 
for prompting this detailed scrutiny of the structure of 
the small-scale magnetic fields and for many stimulating 
discussions of the folding effect and of this work. 
We are also grateful to J.~C.~McWilliams for a number 
of valuable comments. 
We would like to thank the anonymous referees for their 
suggestions, which led to a considerable improvement 
in our presentation. One of the referees also pointed 
out several important 
references~\cite{Pope_curvature,Muzzio_etal,Boozer_etal}. 
The supercomputers used for the simulations quoted in this 
paper are operated by the Caltech Center for Advanced Computing 
Resources and its very helpful staff. 
The work at UCLA was supported by the NSF~Grant~No.~AST~97-13241 
and the DOE~Grant~No.~DE-FG03-93ER54~224.
L.~M.~would like to thank the Department of Astrophysical 
Sciences at Princeton University for financial support.

\appendix

\section{Derivation of the Fokker-Planck Equation for 
the Joint PDF of Magnetic Field and Lorentz Tension}
\label{ap_PDF_derive}

Let us briefly describe the (standard) procedure 
we used to derive the Fokker-Planck equation~\exref{FPEq_BF}. 

In the case of incompressible advecting flow, 
the magnetic field and the Lorentz tension satisfy 
\bea
\label{B_eq_inc}
\dt B^i + \xi^k B^i_{,k} &=& \xi^i_{,k} B^k,\\
\label{F_eq_inc}
\dt F^i + \xi^k F^i_{,k} &=& \xi^i_{,k} F^k + \xi^i_{,km} B^k B^m.
\eea
We start by introducing the {\em characteristic function} 
of the fields~$\vB(t,\vx)$ and~$\vF(t,\vx)$ at an arbitrary 
fixed point~$\vx$, 
\bea
Z(t;\mu,\lambda) = \bl<\tZ(t,\vx;\mu,\lambda)\br> 
= \bl<\exp\bl[i\mu_i B^i(t,\vx) + i\lambda_i F^i(t,\vx)\br]\br>
\eea
Here and in what follows the angular brackets denote ensemble 
averages and overtildes designate unaveraged quantities. 
The function~$Z(t;\mu,\lambda)$ is the Fourier transform of 
the joint~PDF of the vector elements~$B^i(t,\vx)$ and $F^i(t,\vx)$. 
Clearly, $Z$~cannot have any spatial dependence due to the 
homogeneity of the problem.  

Upon taking the time derivative of the unaveraged 
function~$\tZ(t,\vx;\mu,\lambda)$ and making use of 
the evolution equations~\exref{B_eq_inc} and~\exref{F_eq_inc}, 
we find that $\tZ$~satisfies 
\bea
\label{tZ_eq}
\dt\tZ + \xi^k\tZ_{,k} = 
\(\mu_i{\d\over\d\mu_k} + \lambda_i{\d\over\d\lambda_k}\)\xi^i_{,k}\tZ
-i\lambda_i{\d^2\over\d\mu_k\d\mu_m}\,\xi^i_{,km}\tZ.
\eea
In order to establish an evolution equation for 
the (averaged) characteristic function~$Z(t;\mu,\lambda)$, 
we must average the three mixed products of~$\tZ$ and the 
velocity field that appear in the above equation.  
The average that arises from the 
convective term vanishes due to the incompressibility 
of the velocity field and the homogeneity of the problem: 
$\bl<\xi^k\tZ_{,k}\br> = -\bl<\xi^k_{,k}\tZ\br> = 0$. 
The remaining two averages are computed via the standard 
Gaussian splitting mechanism~\cite{Furutsu_Novikov} 
\bea
\label{avg1}
\bl<\xi^i_{,k}\tZ\br> = -\bl<\xi^i\tZ_{,k}\br>
&=& -\int_0^t\diff t'\int\diff^d x'
\bl<\xi^i(t,\vx)\xi^j(t',\vx')\br>{\d\over\d x^k}
\<\delta\tZ(t,\vx)\over\delta\xi^j(t',\vx')\>
= -{1\over2}\,\kappa^{ij}_{,kl}
\(\mu_j{\d\over\d\mu_l} + \lambda_j{\d\over\d\lambda_l}\)Z,\\ 
\label{avg2}
\bl<\xi^i_{,km}\tZ\br> = \bl<\xi^i\tZ_{,km}\br>
&=& \int_0^t\diff t'\int\diff^d x'
\bl<\xi^i(t,\vx)\xi^j(t',\vx')\br>{\d^2\over\d x^k\d x^m}
\<\delta\tZ(t,\vx)\over\delta\xi^j(t',\vx')\>
= -i\,{1\over2}\,\kappa^{ij}_{,kmln}
\lambda_j{\d^2\over\d\mu_l\d\mu_n} Z,
\eea
where we abbreviate~$\kappa^{ij}_{,kl}=\kappa^{ij}_{,kl}(\vy=0)$, 
and~$\kappa^{ij}_{,kmln}=\kappa^{ij}_{,kmln}(\vy=0)$.
The above expressions have been obtained as follows. 
The functional derivative that appears under the integrals 
is the first-order averaged response function. It satisfies 
the causality constraint in that it vanishes for~$t'>t$, whence 
follows the upper limit of the time integrations. Since the velocity 
field~$\xi^i$ is $\delta$~correlated in time, the time integration 
is removed and only the equal-time value of the response 
function has to be calculated. That is done by formally 
integrating~\eqref{tZ_eq} from~$0$ to~$t$, taking the functional 
derivative~$\delta/\delta\xi^j(t',\vx')$ of both sides, 
averaging, setting~$t=t'$, and taking causality into account. 
The result~is 
\bea
\label{resp_fn}
\<\delta\tZ(t,\vx)\over\delta\xi^j(t,\vx')\> = 
\(\mu_j{\d\over\d\mu_l} + \lambda_j{\d\over\d\lambda_l}\)Z\, 
{\d\over\d x^l}\,\delta(\vx-\vx')
-i\lambda_j{\d^2\over\d\mu_l\d\mu_n} Z\,
{\d^2\over\d x^l\d x^n}\,\delta(\vx-\vx').
\eea
After integration by parts, the spatial integrations are removed due 
to the presence of $\delta$~functions. Note that we make use 
of the fact that odd derivatives of the velocity correlation 
tensor~$\kappa^{ij}(\vy)$ vanish at~$\vy=0$. 

Upon averaging both sides of~\eqref{tZ_eq} and 
using the expressions~\exref{avg1} and~\exref{avg2} 
for the mixed averages, 
we obtain a closed evolution equation for the characteristic function~$Z$,
\bea
\dt Z = -{1\over2}\,\kappa^{ij}_{,kl}
\(\mu_i{\d\over\d\mu_k} + \lambda_i{\d\over\d\lambda_k}\)
\(\mu_j{\d\over\d\mu_l} + \lambda_j{\d\over\d\lambda_l}\)Z 
-{1\over2}\,\kappa^{ij}_{,klmn}
\lambda_i\lambda_j{\d^4\over\d\mu_k\d\mu_l\d\mu_m\d\mu_n}\,Z.
\eea 
Inverse Fourier transforming this equation yields the desired 
Fokker-Planck equation~\exref{FPEq_BF} 
for the joint one-point probability density function 
of the magnetic field~$\vB$ and the Lorentz tension~$\vF$.

The derivation of all other Fokker-Planck equations that appear 
in this paper follows the same general outline.

\section{Compressibility Effects} 
\label{ap_compress_effects}

Let us relax the incompressibility condition and allow 
the advecting velocity field to possess an arbitrary degree 
of compressibility. Mathematically this means that 
we have to retain the terms involving divergences of~$\vu$ 
in the equations~\exref{B_eq} and~\exref{F_eq} and to allow 
the compressibility parameters~$a$ and~$b$ in the small-scale 
expansion~\exref{xi_024} of the velocity correlator to 
vary in the intervals 
\bea
-{1\over d+1}\le a \le 1, \qquad -{2\over d+3}\le b \le 2,
\eea
where the lower bounds correspond to the incompressible 
and the upper to the irrotational case. 
We will often use an alternative pair of compressibility parameters 
$\beta=d[1+(d+1)a]$ and $\zeta=d[2+(d+3)b]$ 
that have the advantage of being always nonnegative and 
vanishing in the case of incompressible velocity field.

The exact treatment of the joint probability distribution 
of~$\vF$ and~$\vB$ is completely analogous to that presented 
in~\secref{stats_FB} for the incompressible case. 
The Fokker-Planck equation is now 
\bea
\nonumber
\dt P &=& -{1\over2}\,\kappa^{ij}_{,kl}
\(-\delta^k_i + {\d\over\d B^i}B^k - \delta^k_i{\d\over\d B^r}B^r 
+ {\d\over\d F^i}F^k - 2\delta^k_i{\d\over\d F^r}F^r\)\\
\nonumber
& & \qquad\,\,\,\times\({\d\over\d B^j}B^l - \delta^l_j{\d\over\d B^s}B^s 
+ {\d\over\d F^j}F^l - 2\delta^l_j{\d\over\d F^s}F^s\) P\\ 
& & +\,{1\over2}\,\kappa^{ij}_{,klmn} 
\({\d\over\d F^i}B^k B^m - \delta^k_i{\d\over\d F^r}B^r B^m\)
\({\d\over\d F^j}B^l B^n - \delta^l_j{\d\over\d F^s}B^s B^n\) P.
\label{FPEq_compr}
\eea

The quantities~$\Fsq$ and~$\Bfr$ again satisfy equations~\exref{Fsq_eq} 
and~\exref{Bfr_eq}, respectively, with coefficients~$\gamma_F$, 
$S_F$, and~$\gamma_4$ modified to include the dependence 
on the compressibility parameters~$\beta$ and~$\zeta$. 
The general expressions for these coefficients are listed 
in~\tabref{tab_gammas}. 
Note that the source term~$S_F$ remains positive for 
all allowed values of~$b$. 
The steady-state solution of the form~\exref{ksq_steady} 
continues to exist provided
$\gamma_F - \gamma_4 < 0$, 
which is satisfied for values of the compressibility 
parameter~$a$ such that 
\bea
\label{a_crit}
a < a_c = {d-2\over2(3d-2)} 
\eea
(in $d=2$, $a_c=0$, in $d=3$, $a_c=1/14$; 
this inequality can also be 
derived from a generalization of the simple argument in support of 
folding given in the Introduction: see~\apref{ap_ac_from_metric}).
Thus, for ``nearly incompressible'' flows, the folding picture 
persists in the strong sense that the parallel scale of the field 
remains approximately constant and comparable to the characteristic 
scale of the advecting flow. On the other hand, if the flow possesses 
a fair degree of compressibility, the parallel scales will start 
decreasing exponentially. 

Let us now retrace the path taken in~\secref{stats_FB} and 
study the evolution of mean-square curvature and mirror force 
in the case of arbitrary degree of compressibility. 
Again, equations~\exref{Ksq_eq} and~\exref{Msq_eq} preserve 
their form with modified coefficients~$\gamma_K$, $S_K$, 
$\gamma_M$, $\gamma_{MK}$, $S_M$ (see~\tabref{tab_gammas}). 
None of these quantities changes its sign for any 
allowed values of the compressibility parameters. 
The essential structure of the solutions, therefore, 
does not change compared to the incompressible case, and 
the growing mean-square curvature~$\Ksq$ remains the one interesting 
quantity to watch.

As we discovered from the statistics of the Lorentz tension, 
for~$a<a_c$ the anticorrelation between the magnetic-field 
strength and the field-line curvature is preserved: while
$\Fsq/\Bfr$~remains constant, $\FB$~grows at the rate~$\gamma_K$. 
However, once the compressibility parameter~$a$ exceeds the critical 
value~$a_c$, the ratio $\Fsq/\Bfr$~starts growing as well, 
and the anticorrelation between~$B$ and~$K$ is weakened. 
Comparing the growth rate~$\gamma_K$ of the mean-square curvature 
with the growth rate~$\gamma_F-\gamma_4$ of the ratio~$\Fsq/\Bfr$, 
we find~that $\gamma_K>\gamma_F-\gamma_4$~provided 
\bea
\label{a_star}
a < a_* = {3\over4d-7}.
\eea
While in~2D the second critical value~$a_*=3$ lies outside 
of the interval of allowed values of~$a\in[-1/3,1]$, 
in~3D we have~$a_*=3/5<1$, which is permitted. Thus, in three dimensions, 
for~$a>3/5$, the negative correlation between the field strength 
and the field-line curvature is replaced by a positive one, 
so the regions of maximal growth of the field and its curvature 
coincide! 

To prevent any misconception from arising with regard to the 
quantitative character of the conditions~\exref{a_crit} 
and~\exref{a_star}, we ought to remark here that the particular 
critical values of the compressibility parameter when one or other 
statistical correlation breaks down are, of course, largely 
functions of what particular statistical averages are 
used to measure these correlations. Such sensitivity is due 
to the high degree of intermittency of the statistics of 
passively advected fields. 

Let us discuss the implications of the new facts that have 
emerged from this excursion beyond the confines of the incompressible 
advection theory. Clearly, the main feature of the compressible 
regime is that the velocity field is freed from having to preserve 
the volume and, along with stretching vortical motions that 
characterized the incompressible case, there now are motions 
that contract (or inflate) the volumes, with magnetic-field lines 
trapped inside. The structure of the magnetic field now depends on 
the competition of stretching and contraction, whose relative importance 
is measured by the compressibility parameter~$a$. We have seen 
in this section that stretching wins as long as $a$~stays below 
a certain critical value~$a_c$. Once this value is exceeded, the parallel 
scale of the field cannot be prevented from decaying exponentially. 
While it may still be decaying slower than the perpendicular scale, 
thus giving rise to ``small folds,'' both scales are now deep 
in the subviscous range and will eventually equalize when the resistive 
cut-off scale is reached. A tangled state will result. 
As $a$~increases, the anticorrelation between the strength of 
the field and its curvature gradually weakens and, in~3D, is even 
reversed when $a$~reaches a second critical value~$a_*$. 
This gives another indication of the increasingly tangled 
nature of the growing magnetic field. 

However, as is seen in~\secref{stats_K}, 
the growth of the magnetic field in sufficiently compressible 
flows only takes place in a small fraction of the total 
volume of the system, while elsewhere both the field strength 
and the field-line curvature remain relatively low. 
Thus, the tangled state is not set up everywhere throughout the system, 
but only in a small part of it where there is an appreciable 
growing magnetic field. This situation is, of course, due 
to volume contraction.  
The distribution of the density of the advecting medium 
is lognormal (highly intermittent)~\cite{BS_metric}. 
While~$\langle\rho^2\rangle$ and all higher density moments 
grow exponentially [\eqref{rho_2n}~of~\apref{ap_ac_from_metric}], 
the growth of the density only occurs in a small fraction of 
the volume of the system. This is very natural and could not have 
been otherwise, for, as the total mass of the medium 
is conserved, $\langle\rho\rangle=\const$, the exponential growth 
of the higher moments of the density must be compensated for by the 
exponential contraction of the regions that are responsible for this 
growth. (In this context, one may also recall the results of 
Chertkov~{\em et al.}~\cite{Chertkov_etal_compress} 
who found that for compressible-enough advecting velocity 
fields, the Liapunov exponents for the Lagrangian fluid-particle 
separation become negative, so the fluid-particle trajectories 
tend to converge.) 
The density statistics are known to be intimately related 
to the statistics of the magnetic field (see~Ref.~\cite{BS_metric} 
and~\apref{ap_ac_from_metric}). 
Namely, there is a positive correlation between the density 
of the medium and the strength of the frozen-in magnetic field. 
This positive correlation can be deduced from the fact established 
in~Ref.~\cite{BS_metric} that even moments 
of~$B/\rho^{1-1/d}$ are universal functions independent of 
the density statistics. The magnetic field will, therefore, tend 
to grow wherever the density does. 

Finally, let us note that, in the case of compressible MHD turbulence, 
the condition for the onset of nonlinearity is determined not just 
by the magnitude of the Lorentz tension~$\vB\cdot\nabla\vB$, but rather 
by that of the total Lorentz force divided by the density of the 
fluid: $(-\nabla B^2/2 + \vB\cdot\nabla\vB)/\rho$, 
which includes the magnetic pressure term. 
The latter will grow exponentially due {\em both} to the amplification 
of the magnetic energy {\em and} to the decrease of 
the (total) characteristic scale of the magnetic field. 
It is not hard to see that it will quickly outgrow the tension 
term and the nonlinear effect associated with magnetic pressure 
will dominate. 

\section{An Alternative Derivation of the Critical Degree 
of Compressibility}
\label{ap_ac_from_metric}

Let us demonstrate how the critical value of the compressibility 
parameter~$a$ derived in~\apref{ap_compress_effects} can be 
obtained by 
constructing a generalization of the simple argument we gave 
in the Introduction [formula~\exref{qualit_arg}]. 
Upon using the continuity equation for the density~$\rho(t,\vx)$ 
of the medium, 
\bea
\Dt\rho = -\rho\nabla\cdot\vu,
\eea
we find that the magnetic field~$\vB(t,\vx)$ 
and the Lorentz tension~$\vF(t,\vx)$ satisfy 
\bea
\Dt{\vB\over\rho} &=& {\vB\over\rho}\cdot\nabla\vu,\\
\label{Frho_eq}
\Dt{\vF\over\rho^2} &=& {\vF\over\rho^2}\cdot\nabla\vu 
+ {\vB\over\rho}{\vB\over\rho}:\nabla\nabla\vu 
- {\vB\over\rho}{\vB\over\rho}\cdot\nabla\nabla\cdot\vu.
\eea
We again suppose that the parallel variation of~$\vB$ is initially 
on scales much smaller than those of the velocity field~$\vu$. 
Then the terms in~\eqref{Frho_eq} that contain second-order derivatives 
of~$\vu$ are subdominant and can be neglected. 
We see that in such a case $\vB/\rho$ and $\vF/\rho^2$ 
satisfy the same equation, which is the equation for 
the advection of a (contravariant) passive vector~$\vW$, 
\bea
\Dt\vW = \vW\cdot\nabla\vu.
\eea 
The statistics of passive vectors were treated (as a particular  
case of the statistics of general tensor fields) in~Ref.~\cite{BS_metric}. 
It was proved there that these statistics could be separated 
into two {\em independent} parts: one universal, the other 
nonuniversal, the latter being expressible in terms of the statistics 
of the density. Specifically, the even moments of~$\vW\rho^{1/d}$ 
are universal functions independent of the statistics of the density~$\rho$. 
Therefore, the even moments of~$\vB/\rho^{1-1/d}$ and of~$\vF/\rho^{2-1/d}$ 
should also be independent of the statistics of the density. 
Making use of these results, one can write 
\bea
\label{B_2n}
\langle B^{2n}\rangle &=& 
\biggl<\({B\over\rho}\)^{2n}\rho^{2n}\biggr> = 
\const\,f_d(n,t)\bl<\rho^{2n(d-1)/d}\br>,\\
\label{F_2n}
\langle F^{2n}\rangle &=& 
\biggl<\({F\over\rho^2}\)^{2n}\rho^{4n}\biggr> = 
\const\,f_d(n,t)\bl<\rho^{2n(2d-1)/d}\br>,
\eea
where $f_d(n,t)=\langle (W\rho^{1/d})^{2n}\rangle$~is a universal function. 
Both~$f_d(n,t)$ and 
the moments of the density field were calculated in~Ref.~\cite{BS_metric},
\bea
f_d(n,t) &=& \const\,\exp\[{d-1\over d}\,n(2n+d)(1+a)\kappa_2 t\],\\ 
\label{rho_2n}
\langle\rho^{2n}\rangle &=& \const\,\exp\bl[n(2n-1)\beta\kappa_2 t\br], 
\eea  
where~$\beta = d[1+a(d+1)]$ (vanishes in the incompressible case). 
With the aid of the above formulas, it is straightforward to calculate 
\bea
\ksq = {\Fsq\over\Bfr} \propto e^{\gpar t},\qquad 
\gpar = \bl[2(3d-2)a - (d-2)\br]\kappa_2.
\eea
We see that $\gpar<0$ for values of the compressibility 
parameter~$a$ such that
\bea
a < a_c = {d-2\over2(3d-2)}.
\eea
We have thus recovered the inequality~\exref{a_crit}.

We would like to emphasize that the above derivation clearly 
demonstrates that the effects of compressibility on the field structure 
are due to the crucial part that the density of the medium plays in 
determining the statistics of the magnetic field in compressible flows.


\begin{thebibliography}{99}
\bibitem{fnote_mag_pressure}
In the incompressible MHD turbulence, the magnetic pressure 
term~$-\nabla B^2/2$ need not be considered separately: 
it is combined with the fluid pressure and the total 
pressure is then determined from the incompressibility condition. 
This leads effectively to the tension term~$\vB\cdot\nabla\vB$ 
being acted on by the projection 
operator~$\proj=\unity-\nabla\Delta^{-1}\nabla$, where $\unity$~is 
a unit dyadic. This operator does not change the order of 
magnitude of the tension.
\bibitem{fnote_mat_surf}
The advection of material surfaces is related to 
the passive-scalar-gradient advection just as 
the advection of material lines is related to the kinematic 
dynamo~\cite{Batchelor_mat_lines}. While the curvature 
statistics for line and surface elements exhibit qualitatively 
similar behavior, it would not be surprising to find 
differences on the quantitative level: e.g., different 
power laws for the tails of the curvature~PDF's.
\end{thebibliography}
\end{document}